\documentclass[a4paper,11pt]{article}
\usepackage{jcappub} 
\usepackage{bm}
\usepackage[T1]{fontenc} 

\title{Testing the polarization of gravitational wave background with LISA-TianQin network}

\author[a]{Yu Hu,}
\author[a]{Pan-Pan Wang,}
\author[a]{Yu-Jie Tan}
\author[a,1]{Cheng-Gang Shao \note{Corresponding author.}}

\affiliation[a]{MOE Key Laboratory of Fundamental Physical Quantities Measurement, Hubei Key Laboratory of Gravitation and Quantum Physics, PGMF, and School of Physics, Huazhong University of Science and Technology, 430074, Wuhan, Hubei, China}

\emailAdd{yuhu@hust.edu.cn}
\emailAdd{ppwang@hust.edu.cn}
\emailAdd{yjtan@hust.edu.cn}
\emailAdd{cgshao@hust.edu.cn}

\abstract{While general relativity predicts only two tensor modes for gravitational wave polarization, 
general metric theories of gravity allows up to four additional modes, including two vector and two scalar modes.
Observing the polarization modes of gravitational waves could provide a direct test of the modified gravity.
The stochastic gravitational wave background (SGWB), which may be detected by space-based laser-interferometric detectors at design sensitivity, will provide an opportunity to directly measure alternative polarization.
In this paper, we investigate the performance of the LISA-TianQin network for detecting alternative polarizations of stochastic backgrounds, and propose a method to separate different polarization modes.
First, we generalize the small antenna approximation to compute the overlap reduction functions for SGWB with arbitrary polarization, which is suitable for any  time-delay interferometry combination.
Then we analyze the detection capability of LISA-TianQin for SGWB with different polarizations.
Based on the LISA-TianQin orbital characteristics, we propose a method to distinguish different polarization modes from their mixed data.
Compared with ground-based detectors, the LISA-TianQin network is more capable of resolving polarizations of SGWB.
In particular, the LISA-TianQin network has the potential to resolve two scalar modes that ground-based detectors cannot.
}

\begin{document}
\maketitle
\flushbottom

\section{Introduction}\label{sec1}

Since the first gravitational wave (GW) signal GW150914 was detected by Advanced LIGO in 2015~\cite{2016prl_GW150914},
Advanced LIGO and later Advanced Virgo have detected more and more GWs sourced from compact binary coalescences\cite{2017prl_GW170817,2021apjl_NSBH, 2016prx_LIGO_O1, 2019prx_GWTC-1,2021prx_GWTC-2}.
At the same time, this also implies that there should be many weak GWs that cannot be recognized by detectors.
The combined weak signal from the population of compact binary constitutes the stochastic gravitational wave background (SGWB).
In addition to the astrophysical sources, there are many ways to generate SGWB in the early Universe, 
such as cosmological phase transitions~\cite{1986mnras_GWB_cosmol_pt,2010prd_GWB_QCD_pt}, 
inflation~\cite{1992pr_cosmol_perturbation,1997prd_GWB_inflation, 2021mnrasl_inflationary_SGWB_NANOGrav, 2022prd_inflationary_SGWB_NANOGrav}, 
cosmic strings~\cite{2005prd_GWB_cosmic_strings,2007prl_GWB_cosmic_strings}, etc.
The detection of the GWB can provide us with information on the astronomical distribution~\cite{2014prd_astro_motivation_SGWB, 2016prx_limits_astro_GWB} and cosmology~\cite{2001prd_early_universe_LISA, 2016jcap_cosmol_pt_LISA, 2016jcap_GWB_inflation_LISA,2020jcap_cosmol_pt_LISA, 2020jcap_GWB_CS_LISA}, 
and also provide an opportunity to test the theory of gravity~\cite{2016prl_constrain_MTG_GWB, 2017prx_polarization_SGWB}.

Pulsar Timing Arrays (PTA) has accumulated more than a decade of data, and is expected to detect SGWB in the nHz band in the near future~\cite{2020apjl_GWB_NANOGrav,2021mnras_GWB_EPTA,2021apjl_GWB_PPTA,2022mnras_GWB_IPTA}.
In the Hz hand, the ground-based laser interferometers give an upper limits of fractional energy density of SGWB $\Omega_{\text{GW}}\leq5.8\times 10^{-9}$~\cite{2021prd_upper_limits_GWB_LIGO_O3}.
The future space-based interferometers such as LISA~\cite{2017arX_LISA}, TianQin~\cite{2016cqg_TQ}, Taiji~\cite{2017nsr_Taiji} and DECIGO~\cite{2011cqj_DECIGO}, will be able to detect SGWB with high sensitivity in the mHz band~\cite{2022prd_TQ_GWB, 2021prd_TJ_GWB, 2020prd_mHz_GWB}.
Once we have detected the SGWB, we next need to know what information can be obtained from it.
An important role of SGWB is to test the theory of gravity.
General relativity (GR) predicts that there only exists two tensor polarizations for gravitational wave, the plus and cross modes.
However, there are additional four polarizations allowed by the generic metric theories of gravity, 
including two vector modes and two scalar modes\cite{1961pr_Brans_Dicke, 1973prd_test_polarization_GW, 2004prd_aether_wave, 2010prd_TeVeS_GW, 2017prd_fR_polarization, 2019prd_fR_polarization, 2020epjc_landau_damp_scalar, 2021universe_longit_polarization}.
The observation of vector or scalar modes would cast doubt on general relatively, 
and the absence could also be used to constrain modified gravity~\cite{2014lrr_GR_test}.
If parity is considered, there is circular polarization in the SGWB, describing the asymmetry of the two tensor polarizations~\cite{2007prl_GWB_parity, 2008prd_GWB_parity, 2016prd_circular_SGWB_PTA, 2021prd_circular_SGWB_PTA_orf}.

Research on the detection of polarization modes of the SGWB has become a hot topic, 
as more and more modified theories of gravity are proposed.
For PTA, the detectability of non-GR polarizations of the SGWB was first investigated in~\cite{2008apj_Lee_ORF_nonGR}.
After that, more detailed extension can be found in~\cite{2012prd_ORF_nonGR, 2015prd_PTA_GWB_nonGR}.
No evidence of non-GR polarizations of SGWB has been found in more than a decade of PTA data.
The constrains of the amplitude or energy density for non-GR modes in SGWB at frequency of 1/yr can be found in 
\cite{2018prl_constrain_nonGR_PTA_GWB, 2021apjl_cosntrain_nonGR_PTA_GWB, 2022apj_constrain_nonGR_PTA_GWB}.
On the other hand, the ground-based gravitational-wave detection network is gradually forming, as more and more detectors join in.
It has the potential to detect the SGWB produced by compact binary mergers.
Data from Advanced LIGO's and Advanced Virgo's three observing run can constrain the fractional energy density of different polarization modes at Hz band~\cite{2018_nonGR_GWB_LIGO_O1, 2019prd_LIGO_O2_GWB, 2021_GWB_O3}.
However, the ability of the ground-based laser interferometer network to study the polarizations of SGWB is limited.
The current sensitivity of ground-based detectors is limited enough to detect the SGWB.
What more, the two kind of scalar modes are completely different, one is the transverse and the other is longitudinal.
Unfortunately, the responses of the ground-based laser interferometer to scalar-breathing and scalar-longitudinal modes are completely degenerate, which means that the two modes cannot be distinguished, no matter how sensitive the detectors are~\cite{2009prd_nonGR_GWB_LIGO}.
However, the two scalar polarizations are of a completely different nature, and they cannot be converted into one another via a
rotational coordinate transformation along the direction of propagation (whereas two tensor modes or two vector modes can).
The physical images behind the appearance of scalar breathing and longitudinal modes are distinct.
For example, in the scalar-tensor theory of gravity, there exists only one extra breathing mode if the scalar field is massless, or a mixture of breathing mode and longitudinal mode if the scalar field is massive~\cite{2017prd_fR_polarizations}.
It is therefore necessary to distinguish between the two scalar modes in the detector.

For the future space-based gravitational-wave detectors, the situation is different.
The degeneracy of response between the breathing and longitudinal modes is broken at relatively high frequencies~\cite{2019prd_response_space_interfer, 2020prd_Taiji_polarization}, which implies that the space-based detectors have potential to distinguish breathing and longitudinal modes of gravitational waves.
Besides, space-based detectors can detect signals from a much larger range of frequencies, compared to ground-based detectors.
The abundance of sources in the sensitive band and the sufficiently high sensitivity of the detectors ensure that the polarization analysis can be performed.
What more,  the relative orientation of two space-based detectors may change as they are in different positions in their respective orbits.
For example, for the LISA-TianQin network, the normal vector of the constellation plane of LISA is vary with time and that of TianQin points to a fixed direction~\cite{2016cqg_TQ}.
This means that the overlap reduction function (ORF), the transfer function between the spectrum of SGWB and the power spectrum of cross correlation signal, will vary accordingly.
In this way, it is easier to distinguish between different polarization modes, 
which can be inferred from the viewpoint that the detectors can be regarded as different detectors at different positions.

For the space-based detectors, the laser phase noise is usually orders of magnitude higher than other noises due to the mismatch of arm lengths, and also much larger than the gravitational wave signal.
Fortunately, the time delay interferometry (TDI) can be used to suppress the laser phase noise~\cite{2000prd_TDI_early, 2002prd_TDI, 2002prd_LISA_optimal_sensitivity, 2020lrr_TDI}.
It is worth mentioning that three noise quadrature channels can be constructed, called A, E and T.
The T channel is relatively insensitive to the gravitational-wave compared with other channels.
So the readout of T channel can be used to model the noise in order to subtract the noise from the A and E channels to get the SGWB.
This is called the null channel method~\cite{2001prd_SGWB_null, 2010prd_SGWB_null, 2017lrr_detection_GWB}. 
However, the persuasiveness of the null channel is limit, it is required that the cross-correlation signals are detected in the data of two detectors to claim detection of SGWB.
The cross-correlation analysis based on TDI combination is a standard method for space-based detectors to detect the SGWB.
However, there is a lack of study on the performance of the space-based detectors for SGWB with different polarizations, as well as the treatment of the time varying effect.
Among the promising space-based detectors, the difference in the orbital configuration of LISA and TianQin is most significant, so the time-varying effect of LISA-TianQin network is the most pronounced. 
In this paper, we investigate the detection capability of the LISA-TianQin network for SGWB with different polarization, and propose a method inspired by the time-varying effect to distinguish different polarization modes from their mixed data.

The outline of the paper is as follows.
In section~\ref{sec2}, we review the SGWB in general metric theories of gravity and introduce the LISA-TianQin network.
In section~\ref{sec3}, we review the correlation analysis for detecting SGWB and calculate the overlap reduction function for LISA-TianQin network.
In section~\ref{sec4}, we study the sensitivity and detectability for the SGWB of alternative polarizations.
Then, in section~\ref{sec5}, we consider the method to separation the polarizations for LISA-TianQin network.
Finally, a discussion is presented in in section~\ref{sec6}.

\section{SGWB statistic and LISA-TianQin network}\label{sec2}
\subsection{Stochastic background of non-GR polarizations}
The metric perturbations corresponding to SGWB can be expressed as a superposition of plane waves of different frequencies from different directions~\cite{2017lrr_detection_GWB}:
\begin{equation}\label{h_ab_t}
    h_{ab}(t,\vec{x})=\int_{-\infty}^{\infty}df \int d^2\Omega_{\hat{n}} h_{ab}(f,\hat{n}) e^{i2\pi f(t+\hat{n} \cdot \vec{x}/c)}.
\end{equation}
The fourier coefficients $h_{ab}(f,\hat{n})$ are random variables, whose statistic is significant.
In generic metric theory, the coefficients can be expanded in terms of the six spin-2 polarization tensors:
\begin{equation}
    h_{ab}(f,\hat{n}) = \sum_A h_{A}(f,\hat{n}) e^{A}_{ab}(\hat{n}) ,
\end{equation}
where $A=\{+, \times, X, Y, B, L\}$ represents different polarization modes, where ${+,\times}$ represent tensor modes predicted by general relativity, and ${X,Y}$ and ${B,L}$ represent vector and scalar modes allowed by the general metric theory of gravity.
Explicitly, the six spin-2 polarization tensors are 
\begin{equation}\label{e_ab_n}
    \begin{aligned}
    e^{+}_{ab}(\hat{n})=&\hat{\theta}_a \hat{\theta}_b - \hat{\phi}_a \hat{\phi}_b , \quad
    e^{\times}_{ab}(\hat{n})=\hat{\theta}_a \hat{\phi}_b + \hat{\phi}_a \hat{\theta}_b ,\\
    e^X_{ab}(\hat{n})=&\hat{\theta}_a\hat{n}_b+\hat{n}_a\hat{\theta}_b ,\quad
    e^Y_{ab}(\hat{n})=\hat{\phi}_a\hat{n}_b+\hat{n}_a\hat{\phi}_b ,\\
    e^B_{ab}(\hat{n})=&\hat{\theta}_a\hat{\theta}_b+\hat{\phi}_a\hat{\phi}_b ,\quad
    e^L_{ab}(\hat{n})=\sqrt{2}\hat{n}_a\hat{n}_b ,
\end{aligned}
\end{equation}
and $\hat{\theta}$, $\hat{\phi}$ are the standard angular unit vectors tangent to the sphere:
\begin{equation}\label{n}
    \begin{aligned}
        \hat{n}     &=(\sin\theta \cos\phi,\sin\theta \sin\phi,\cos\theta),\\
        \hat{\theta}&=(\cos\theta \cos\phi,\cos\theta \sin\phi,-\sin\theta) ,\\
        \hat{\phi}  &=(-\sin\phi,\cos\phi,0) .
    \end{aligned}
\end{equation}

The statistical properties of the SGWB are described by the probability distribution of the metric perturbations.
In this work, we assume that the SGWB is Gaussian, stationary and isotropic.
And without loss of generality we can assume that the background has zero mean $\left\langle h_{A}(f, \hat{n}) \right\rangle=0$.
So all the information is encoded in the quadratic expectation:
\begin{equation}\label{S_h}
    \left\langle h_{A}(f, \hat{n}) h_{A^{\prime}}^{*}\left(f^{\prime}, \hat{n}^{\prime}\right)\right\rangle
    =\frac{1}{8 \pi} S^A_{h}(f) \delta\left(f-f^{\prime}\right) \delta_{A A^{\prime}} \delta^{2}\left(\hat{n}, \hat{n}^{\prime}\right) ,
\end{equation}
where $S^A_{h}(f)$ can be regarded as the component corresponding to the $A$ polarization of a one-sided gravitational-wave strain power spectral density function.
We further assume that both the tensor and vector modes are unpolarized, 
which implies that
\begin{equation}
    \begin{aligned}
    S^{+}_{h}&=S^{\times}_{h}=S^{T}_{h}/2, \\
    S^{X}_{h}&=S^{Y}_{h}=S^{V}_{h}/2.
    \end{aligned}
\end{equation}
However, the two scalar modes should be considered as two independent polarization modes, 
since one is the longitudinal and the other is transverse.
It is worth noting that some models predict the presence of parity violating SGWB sources \cite{2006prl_polarized_GWB_inflation, 2011jcap_parity_violation_CMB_inflaton}, which can be recognized with the interferometers \cite{2006prl_circular_GWB_detection, 2020jcap_circular_GWB, 2021jcap_circular_GWB_LISA_Taiji}.

The function $S^A_{h}(f)$, which characterizes the spectral shape of the SGWB within each polarization sector, 
can be detected directly without assuming a model.
However, the amplitude of SGWB for each polarization is characterized by the fractional energy density~\cite{1999prd_SGWB_detecte},
\begin{equation}\label{Omega}
    \Omega^A_{\mbox{gw}}(f)=\frac{1}{\rho_{c}} \frac{d \rho^A_{\mbox{gw}}}{d \ln f} ,
\end{equation}
defined as the energy density per logarithmic frequency bin, normalized by the critical energy density of the closed Universe $\rho_{c} \equiv 3 c^{2} H_{0}^{2} / 8 \pi G $.
Here $G$ is the gravitational constant, and $H_0=67.4\text{km s}^{-1} \text{Mpc}^{-1}$ is the Hubble constant~\cite{2020aa_Planck2018_cosmol}.
In general relatively, the relation between $S^A_{h}(f)$ and $\Omega^A_{\mbox{gw}}(f)$ is~\cite{1999prd_SGWB_detecte}
\begin{equation}\label{OtoS}
    \Omega^A_{\mbox{gw}}(f)=\frac{2\pi^2}{3H_0^2} f^3 S^A_{h}(f).
\end{equation}
In alternative theories of gravity, eq.~(\ref{OtoS}) may not hold unless the stress-energy of gravitational waves also obeys Isaacson's formula~\cite{1968prd_Isaacson_Tuv_GW}:
\begin{equation}
    \rho_{\mbox{gw}} = \frac{c^2}{32\pi G} \left\langle \dot{h}_{ab}(t,\vec{x}) \dot{h}^{ab}(t,\vec{x}) \right\rangle .
\end{equation}
In this case, $\Omega^A_{\mbox{gw}}(f)$ can be understood as a function of the observable $S^A_{h}(f)$ rather than the fractional energy density.

Many theoretical models of SGWB predict that the shape of $\Omega^A_{\mbox{gw}}(f)$ can be modeled as power laws~\cite{2017lrr_detection_GWB}, such that
\begin{equation}
    \Omega^A_{\mbox{gw}}(f)=\Omega^{\alpha_A}_{0}\left(\frac{f}{f_0}\right)^{\alpha_A}.
\end{equation}
Here $\Omega^{\alpha_A}_{0}$ is the amplitude of polarization $A$ at a reference frequency $f_0$ and $\alpha_A$ is the corresponding spectral index.
For instance, the tensor polarization background from compact binary coalescences is modeled by power law with index $\alpha_T=2/3$~\cite{2019prd_LIGO_O2_GWB} and for the inflationary cosmic background is $\alpha_T=0$~\cite{2001cqg_cosmic_SGWB_space_detect}.

\subsection{LISA-TianQin network}
The LISA and TianQin are proposed space GW missions targeting to detect the GW in the frequency band 0.1 mHz -- 1 Hz.
The difference is that LISA is heliocentric orbit and TianQin is geocentric.
In addition, the relative angle between their detecter plane will vary with time.
TianQin has a geocentric orbit and consists of three satellite formations in a nearly equilateral triangle.
Accurate to the first order, the coordinates of the three satellites of TianQin are~\cite{2018cqg_TQ_orbit}
\begin{equation}
    \vec{r}^{TQ}_{n} = \vec{r}^{TQ}_0 + \vec{R}^{TQ}_n ,
\end{equation}
where $\vec{r}^{TQ}_0 = (x^{TQ}_0, y^{TQ}_0, z^{TQ}_0)$ is the geocentric coordinate,
\begin{equation}
    \begin{aligned}
        x^{TQ}_0 &= R\cos\alpha_{TQ} +\frac{1}{2}e^{TQ}R(\cos 2\alpha_{TQ}-3), \\
        y^{TQ}_0 &= R\sin\alpha_{TQ} +\frac{1}{2}e^{TQ}R\sin 2\alpha_{TQ}, \\
        z^{TQ}_0 &= 0 ,
    \end{aligned}
\end{equation}
and $\vec{R}^{TQ}_n = (X^{TQ}_n, Y^{TQ}_n, Z^{TQ}_n)$ are the coordinates of the satellites in the geocentric coordinate system, 
\begin{equation}
    \begin{aligned}
        X^{TQ}_n &= R_{1}\left(\cos \phi_{s} \sin \theta_{s} \sin \alpha_{n}+\cos \alpha_{n} \sin \phi_{s}\right),  \\
        Y^{TQ}_n &= R_{1}\left(\sin \phi_{s} \sin \theta_{s} \sin \alpha_{n}-\cos \alpha_{n} \cos \phi_{s}\right),  \\
        Z^{TQ}_n &= -R_{1} \sin \alpha_{n} \cos \alpha_{n} . 
    \end{aligned}
\end{equation}
Here $R = 1\text{AU}$, $e^{TQ} = 0.0167$, $\alpha_{TQ} = 2\pi f_m t - \alpha_0$, $f_m = 1/\text{yr}$, $\alpha_0 = 102.9^{\circ}$,
$\alpha_{n} = 2\pi f_{sc}t + \kappa_n$, $\kappa_n = 2/3(n-1)\pi$, $R_1 = 1 \times 10^8 m $, $\theta_{s}=-4.7^{\circ}$, 
$\phi_{s}=120.5^{\circ}$, $f_{sc} = 1/(3.64\text{days})$ and $n=1,2,3$ respectively denotes one of the three satellites.
The detector plane orientation is fixed as $(\cos\theta_s\cos\phi_s, \cos\theta_s\sin\phi_s, \sin\theta_s)$ and the arm length is $L^{TQ}=\sqrt{3} \times 10^8 \text{m} $.
The displacement measurement noise $\sqrt{S^{TQ}_x}=1 \times 10^{-12}\text{m}/\text{Hz}^{-1/2}$ and the residual acceleration
noise $\sqrt{S^{TQ}_a}=1 \times 10^{-15}\text{m s}^{-2}/\text{Hz}^{-1/2}$.
LISA has a heliocentric orbit at $20^\circ$ behind the Earth.
The satellite formation consists of three satellites to form an approximate equilateral triangle~\cite{2005cqg_LISA_orbit}, and the coordinates of the three LISA satellites are $r^{LS}_n = (x^{LS}_n, y^{LS}_n, z^{LS}_n)$, namely,
\begin{equation}
    \begin{aligned}
        x^{LS}_n &=  R\cos \alpha_{L S}+\frac{1}{2} e^{LS}R\left(\cos \left(2 \alpha_{L S}-\kappa_{n}\right)-3 \cos \kappa_{n}\right), \\
        y^{LS}_n &=  R\sin \alpha_{L S}+\frac{1}{2} e^{LS}R\left(\sin \left(2 \alpha_{L S}-\kappa_{n}\right)-3 \sin \kappa_{n}\right), \\
        z^{LS}_n &= -\sqrt{3} e^{LS} R \cos(\alpha_{LS}-\kappa_{n}),
    \end{aligned}
\end{equation}
where $\alpha_{LS} = \alpha_{TQ} - 20^{\circ} $ and $e^{LS} = 0.0048$.
The detector plane is inclined to the orbit plane by $60^\circ$ and  the arm length is $L^{LS}= 2.5 \times 10^{9} \text{m}$.
The displacement measurement noise $\sqrt{S^{LS}_x}=1.5 \times 10^{-11}\text{m}/\text{Hz}^{-1/2}$ and the residual acceleration
noise $\sqrt{S^{LS}_a}=3 \times 10^{-15}\text{m s}^{-2}/\text{Hz}^{-1/2}$.

\subsection{Noise and response for time delay interferometry}
\begin{figure}[!t]
    \centering
    \includegraphics[width=0.6\textwidth]{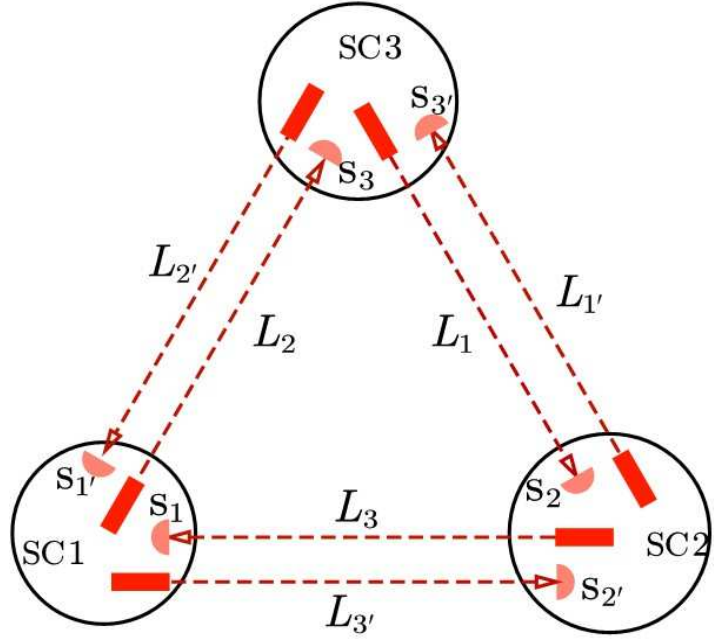} 
    \caption{The configuration of the space-based laser interferometry consists of laser sources and links.}\label{fig1_TDI}
\end{figure}
There are six laser links between the three satellites.
Laser noise can be effectively reduced by constructing time delay interferometry combinations.
Any TDI combination can be expressed in terms of a polynomial of the delay operator acting on the six received signals,
\begin{equation}\label{tdi}
    s_{TDI} = \sum_{i} P_i s_i .
\end{equation}
Here $i = 1,2,3,1^{\prime},2^{\prime},3^{\prime}$ represents a link respectively: $2\rightarrow1$, $3\rightarrow2$, $1\rightarrow3$, $3\rightarrow1$, $1\rightarrow2$, $2\rightarrow3$.
The detector configuration is illustrated in figure~\ref{fig1_TDI}.
The time delay operator is defined as $D_i s_j(t) = s_j(t-L_i/c)$, 
and converted to the frequency domain to $\tilde{D}_i  = e^{-i2\pi fL_i/c}$.
For example, the coefficients of the first-generation TDI Michelson combination $X$ are given by 
\begin{equation}
    \begin{aligned}
        P_1 &= D_{2^{\prime}2}-1 , P_2 = 0, P_3 = D_{2^{\prime}}-D_{33^{\prime}2^{\prime}} \\
        P_{1^{\prime}} &= 1-D_{33^{\prime}} , P_{2^{\prime}} = D_{2^{\prime}23}-D_{3}, P_{3^{\prime}} = 0, \\
    \end{aligned}
\end{equation}
which represents a laser interferometry link:
$$ [1\rightarrow 2 \rightarrow 1 \rightarrow 3 \rightarrow 1] - [1\rightarrow 3 \rightarrow 1 \rightarrow 2 \rightarrow 1].
$$
The equivalent expressions for Michelson channel $Y$ and $Z$ can be obtained by permuting the label $\{1,2,3\}$.
Three noise-orthogonal channels can be constrained by the linear combinations of $X, Y, Z$,
\begin{equation}
    A=\frac{Z-X}{\sqrt{2}},E=\frac{X-2Y+Z}{\sqrt{6}},T=\frac{X+Y+Z}{\sqrt{3}}.
\end{equation}

\begin{figure}[!t]
    \centering
    \includegraphics[width=0.6\textwidth]{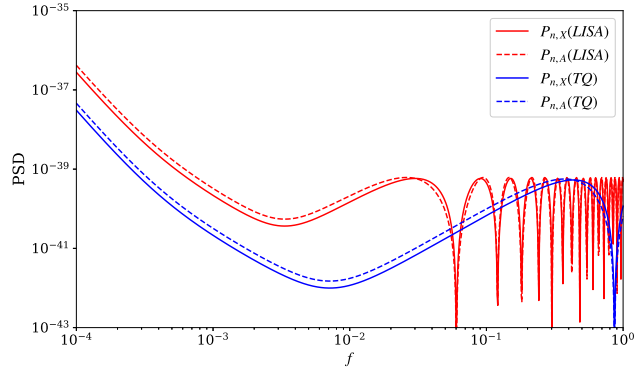} 
    \caption{The PSD $P_n(f)$ for different channels.
    The solid and dashed lines denote the X and A channels respectively. 
    And we adopt the red/blue to label the LISA/TianQin. }\label{fig2_PSD}
\end{figure}

After removing the phase noise, there are residual acceleration noise and displacement measurement noise in the TDI combination.
The power spectral densities (PSDs) of the remaining noises is
\begin{equation}
    \begin{aligned}
        P_{n,TDI} = &\frac{1}{L^2}\left[C_1 S_x \left(1+(2\text{mHz}/f)^4\right) \right. \\
         &\left.+ (2C_1 + C_2\cos\beta)\frac{S_a \left(1+(0.4\text{mHz}/f)^2\right) \left(1+(f/8\text{mHz})^2\right)}{(2\pi f)^4}\right],
    \end{aligned}
    \end{equation}
where $\beta = 2\pi f L/c$ and the coefficients
\begin{equation}
    \begin{aligned}
        C_1 &= \sum_i |\tilde{P}_i|^2 , \\
        C_2 &= \Re(\tilde{P}_1 \tilde{P}^*_{2^{\prime}} + \tilde{P}_2 \tilde{P}^*_{3^{\prime}} + \tilde{P}_3 \tilde{P}^*_{1^{\prime}}) .
    \end{aligned}
\end{equation}
For example, the PSD of $X$ channel is 
\begin{equation}
    \begin{aligned}
        P_{n,X} =& \frac{16\sin^2\beta}{L^2}\left[S_x \left(1+(2\text{mHz}/f)^4\right) \right. \\
        & \left. + \frac{S_a \left(1+(0.4\text{mHz}/f)^2\right) \left(1+(f/8\text{mHz})^2\right)}{(2\pi f)^4}(3 + \cos 2\beta)\right].
    \end{aligned}
    \end{equation}
The PSDs of X and A channel for LISA and TianQin are shown in figure~\ref{fig2_PSD}.

Since the gravitational waves are weak, it is accurate enough to calculate the detecter response to the linear order.
In frequency domain, the gravitational waves signal can be expressed as~\cite{2017lrr_detection_GWB}
\begin{equation}\label{h_f}
    \tilde{h}(f)=\int d^2\Omega_{\hat{n}} \sum_A F^A(f,\hat{n})h_A(f,\hat{n}),
\end{equation}
the response function $F^A(f,\hat{n})= R^{ab}(f,\hat{n})e^A_{ab}(\hat{n})$.
The response for any TDI channel (\ref{tdi}) is
\begin{equation}
    R_{T D I}^{a b}=\sum_{i} P_{i} R^{a b}\left(f, \hat{n}, \hat{u}_{i}, \vec{r}_{i}\right) ,
\end{equation}
where $R^{a b}\left(f, \hat{n}, \hat{u}_{i}, \vec{r}_{i}\right)$ is the impulse response of a single arm, 
$\hat{u}_{i}$ is the direction unit vector of the arm and $\vec{r}_{i}$ is the midpoint of arm.
Here the choose of $\vec{r}_{i}$ is different from the literature for the convenience of calculation,
such that the impulse response become~\cite{1975grg_response_GW}
\begin{equation}
    R^{a b}(f, \hat{n}, \hat{u}, \vec{r})=\frac{1}{2} u^{a} u^{b} \mathcal{T}_{\hat{u}}(f, \hat{n} \cdot \hat{u}) e^{i 2\pi f \hat{n} \cdot \vec{r} / c} ,
\end{equation}
where 
\begin{equation}
    \begin{aligned} \mathcal{T}(f, \hat{n} \cdot \hat{u}) & \equiv \frac{c}{i 2 \pi f L} \frac{1}{1+\hat{n} \cdot \hat{u}}\left[e^{\frac{i \pi f L}{c}(\hat{n} \cdot \hat{u})}-e^{-\frac{i \pi f L}{c}(2+\hat{n} \cdot \hat{u})}\right] \\ &=e^{-\frac{i \pi f L}{c}} \operatorname{sinc}\left(\frac{\pi f L}{c}[1+\hat{n} \cdot \hat{u}]\right) 
    \end{aligned} 
\end{equation}
is the transfer function.

\section{Correlation analysis}\label{sec3}
\subsection{Cross-correlation signal}
Usually, the SGWB is very weak, which is masked by the noise of the detector, and the characteristics are close to the noise. 
So it is difficult to distinguish the noise and the GWB signal in a single detector.
The correlation analysis is a powerful method to detect SGWB~\cite{1999prd_SGWB_detecte}.
We review this method and apply it to the LISA-TianQin network in this section.

We start with the output signals of the two detectors,
\begin{equation}
    \begin{aligned}
        s_I(t) = h_I(t) + n_I(t) ,\\
        s_J(t) = h_J(t) + n_J(t) ,
    \end{aligned}
\end{equation}
where $I,J$ respect TianQin and LISA in this paper.
And the correlation signal is 
\begin{equation}\label{S_t}
    S_{IJ}=\int_{-T / 2}^{T / 2} d t s_{I}(t) s_{J}(t) ,
\end{equation}
or in frequency domain
\begin{equation}\label{S_f}
    S_{IJ}=\int_{-\infty}^{\infty} d f \int_{-\infty}^{\infty} d f^{\prime} \delta_{T}\left(f-f^{\prime}\right) \tilde{s}_{I}(f) \tilde{s}_{J}^{*}\left(f^{\prime}\right) ,
\end{equation}
where $\delta_{T}(f)=\int_{-T / 2}^{T / 2} d t e^{-i 2 \pi f t}=\frac{\sin (\pi f T)}{\pi f} $.
Assume that the two detector noises are uncorrelated, the mean of the correlation signal is
\begin{equation}
    \mu=\langle S_{IJ}\rangle=\int_{-\infty}^{\infty} d f \int_{-\infty}^{\infty} d f^{\prime} \delta_{T}\left(f-f^{\prime}\right)\left\langle\tilde{h}_{I}(f) \tilde{h}_{J}^{*}\left(f^{\prime}\right)\right\rangle .
\end{equation}
Combining eq.~(\ref{S_h}) and eq.~(\ref{h_f}),
\begin{equation}
    \left\langle\tilde{h}_{I}(f) \tilde{h}_{J}^{*}\left(f^{\prime}\right)\right\rangle=\frac{1}{2} \delta\left(f-f^{\prime}\right) \sum_{A=\{T, V, B, L\}} \Gamma_{I J}^{A}(f) S_{h}^{A}(f) ,
\end{equation}
where the overlap reduction functions are
\begin{equation}
    \begin{aligned}
    \Gamma^T_{IJ}(f)=&\frac{1}{8\pi}\int d^2 \Omega_{\hat{n}}\sum_{A={+,\times}}F^{A}_I(f,\hat{n})F^{A*}_J(f,\hat{n}) ,\\
    \Gamma^V_{IJ}(f)=&\frac{1}{8\pi}\int d^2 \Omega_{\hat{n}}\sum_{A={X,Y}}F^{A}_I(f,\hat{n})F^{A*}_J(f,\hat{n}) ,\\
    \Gamma^B_{IJ}(f)=&\frac{1}{4\pi}\int d^2 \Omega_{\hat{n}} F^{B}_I(f,\hat{n})F^{B*}_J(f,\hat{n}) ,\\
    \Gamma^L_{IJ}(f)=&\frac{1}{4\pi}\int d^2 \Omega_{\hat{n}} F^{L}_I(f,\hat{n})F^{L*}_J(f,\hat{n}) .
    \end{aligned}
\end{equation}
We will calculate the ORFs in next subsection.
So the mean of signal is
\begin{equation}
    \mu=\frac{T}{2} \int_{-\infty}^{\infty} d f \sum_{A} \Gamma_{IJ}^{A}(f) S_{h}^{A}(f) .
\end{equation}
Assume that the signal is much smaller than the noise, such that
\begin{equation}
    \left\langle\tilde{s}_{i}^{*}(f) s_{i}^{*}\left(f^{\prime}\right)\right\rangle 
    \approx\left\langle\tilde{n}_{i}^{*}(f) n_{i}^{*}\left(f^{\prime}\right)\right\rangle
    =\frac{1}{2} \delta\left(f-f^{\prime}\right) P_{i}(f) .
\end{equation}
The variance is
\begin{equation}
    \sigma^{2}=\left\langle S_{IJ}^{2}\right\rangle-\langle S_{IJ}\rangle^{2}
    \approx\frac{T}{4} \int_{-\infty}^{\infty} d f P_{I}(f) P_{J}(f) .
\end{equation}
So the signal-to-noise ratio (SNR) is 
\begin{equation}
    \rho=\frac{\mu}{\sigma}=\sqrt{T} 
    \frac{\int_{-\infty}^{\infty} d f \sum_{A} \Gamma_{IJ}^{A}(f) S_{h}^{A}(f)}
    {\sqrt{\int_{-\infty}^{\infty} d f P_{I}(f) P_{J}(f)}} .
\end{equation}

A prerequisite for the cross-correlation analysis is that the ORFs remain constant for the duration of observation.
LISA-TianQin does not meet this condition, so some improvements need to be made.
The total observation time $T$ is divided into $N$ segments, with each segment $\Delta T$.
$\Delta T$ need to be greater than the light travel time between the two detectors and less than the time scale over which the ORF will vary.
Therefore, ORFs in each segment can be treated as constants, labeled as $\Gamma_{IJ,k}^{A}(f)$, and the cross-correlation analysis could be performance.
The SNR for each segment is 
\begin{equation}
    \rho_k=\sqrt{\Delta T} 
    \frac{\int_{-\infty}^{\infty} d f \sum_{A} \Gamma_{IJ,k}^{A}(f) S_{h}^{A}(f)}
    {\sqrt{\int_{-\infty}^{\infty} d f P_{I}(f) P_{J}(f)}} ,
\end{equation}
and the total SNR for observation time $T$ reads
\begin{equation}
    \rho_{total}=\sqrt{\sum_{k=1}^N \rho_{k}^2}.
\end{equation}

\subsection{Overlap reduction function}

The overlap reduction function can be interpretation as the response of correlation of two detectors to the isotropic SGWB~\cite{2017lrr_detection_GWB}.
The overlap reduction function mainly depends on three factors: detector similarity, separation and orientation relative to one another.
In the past literature, one usually considers two identical detectors placed in difference location.
In fact, for detectors like LISA-TianQin, the different arm lengths result in slightly different frequency bands for their respective responses, which is an important factor leading to the reduction of ORF.
On the other hand, changes in orbits cause ORFs to change over time.
In addition, the small-antenna limit that applies to ground-based detectors is no longer always applicable to space-based detectors in the detection band.
For example, the characteristic frequencies $f= c/(2L)$ for LISA and TianQin are $0.06\text{Hz}$ and $0.86\text{Hz}$ respectively.
the small-antenna limit required that the frequency much smaller than the characteristic frequencies.
However, the most sensitive frequency band of LISA-TianQin is $10^{-3}$Hz -- $10^{-1}$ Hz, so the small antenna limit is not always satisfied.
Based on the above considerations, the ORF of LISA-TianQin network deserves a careful discussion.

For any TDI channel, the ORF for $A$ polarization is 
\begin{equation}\label{Gamma_TDI}
    \begin{aligned}
        \Gamma_{TDI}^{A}(f)&=\frac{1}{8 \pi} \int d^{2} \Omega_{\hat{n}} 
    \sum_{A} F_{I}^{A}(f, \hat{n}) F_{J}^{A^{*}}(f, \hat{n}) \\
    &= \frac{1}{8 \pi} \sum_{i, j} P_{I, i} P_{J, j}^{*} \int d^{2} \Omega_{\hat{n}} \sum_{A} R_{I}^{a b}\left(f, n, \hat{u}_{I, i}, \vec{r}_{I, i}\right) R_{J}^{c d^{*}}\left(f, n, \hat{u}_{J, j}, \vec{r}_{J, j}\right) e_{a b}^{A} e_{c d}^{A} \\
    &= \frac{e^{\frac{i}{2}\left(\beta_{J}-\beta_I\right)}}{32 \pi} \sum_{i, j} P_{I, i} P_{J, j}^{*} \hat{u}_{I, i}^{a} \hat{u}_{I, i}^{b} \hat{u}_{J, j}^{c} \hat{u}_{J, j}^{ d} \Gamma^A_{a b c d}\left(\alpha_{i j}, \beta_I, \beta_J, \hat{u}_{I, i}, \hat{u}_{J, j}, s_{i j}\right) ,
    \end{aligned}
\end{equation}
where $\beta_I = 2\pi f L_I/c$ , $\beta_J = 2\pi f L_J/c$ and
\begin{equation}\label{Gamma_abcd}
    \begin{aligned}
    \Gamma^A_{a b c d}&\left(\alpha_{ij}, \beta_I, \beta_J, \hat{u}_{I,i}, \hat{u}_{J,j}, \hat{s}_{i j}\right)=\int d^{2} \Omega_{\hat{n}} e^{-i \alpha_{i j} \hat{n} \cdot \hat{s}_{ij}} \\
    &\times \sum_{A} \operatorname{sinc}\left(\frac{\beta_I}{2}\left[1+\hat{n} \cdot \hat{u}_{I,i}\right]\right) \operatorname{sinc}\left(\frac{\beta_J}{2}\left[1+\hat{n} \cdot \hat{u}_{J,j}\right]\right) 
    e_{a b}^{A} e_{c d}^{A}. 
    \end{aligned}
\end{equation}
Here $ \alpha_{i j} = 2 \pi f s_{i j} / c $, 
$ s_{i j} \equiv\left|\Delta \vec{x}_{i j}\right|=\left|\vec{r}_{J,j}^{\prime}-\vec{r}_{I,i}\right| $
and $ \hat{s}_{i j} = \Delta \vec{x}_{i j} / s_{i j} $.
To keep the definition consistent, the sum means $e_{a b}^{+} e_{c d}^{+} + e_{a b}^{\times} e_{c d}^{\times}$ for $A=T$,
$e_{a b}^{X} e_{c d}^{X} + e_{a b}^{Y} e_{c d}^{Y}$ for $A=V$,
$2e_{a b}^{B} e_{c d}^{B}$ for $A=B$ and $2e_{a b}^{L} e_{c d}^{L}$ for $A=L$.
In this way, the ORF of any TDI channel can be disassembled and calculated between two separate arms.
In general, the integral in eq.~(\ref{Gamma_abcd}) can not be calculated analytically.
In the short antenna limit $\beta,\beta^{\prime} \ll 1$, it can be calculated analytically~\cite{1993prd_orf_LIGO, 1999prd_SGWB_detecte}.
The result is 
\begin{equation}
    \begin{aligned} 
        \Gamma_{a b c d}^{A(0)}(\alpha, \hat{s}) &= A^{A(0)}(\alpha) \delta_{a b} \delta_{c d}
        +B^{A(0)}(\alpha)\left(\delta_{a c} \delta_{b d}+\delta_{b c} \delta_{a d}\right) \\
        &+C^{A(0)}(\alpha)\left(\delta_{a b} s_{c} s_{d}+\delta_{c d} s_{a} s_{b}\right) \\
        &+D^{A(0)}(\alpha)\left(\delta_{a c} s_{b} s_{d}+\delta_{a d} s_{b} s_{c} \right. \\
        &\left.+\delta_{b c} s_{a} s_{d}+\delta_{b d} s_{a} s_{c}\right)
        +E^{A(0)}(\alpha) s_{a} s_{b} s_{c} s_{d} ,
    \end{aligned} 
\end{equation}
where the coefficients is 
\begin{equation}
    X^{A(0)} =(M^{(0)})^{-1}Y^{A(0)}.
\end{equation}
Here,
\begin{equation}
    X^{A(0)} = \left[\begin{array}{c}A^{A(0)} \\ B^{A(0)} \\ C^{A(0)} \\ 
        D^{A(0)} \\ E^{A(0)}\end{array}\right],
    M^{(0)}=\left[\begin{array}{rrrrr}9 & 6 & 6 & 4 & 1 \\
        6 & 24 & 4 & 16 & 2 \\
        6 & 4 & 8 & 8 & 2 \\
        4 & 16 & 8 & 24 & 4 \\
        1 & 2 & 2 & 4 & 1 \\ \end{array}\right],
\end{equation}
and 
\begin{equation}
    Y^{T(0)}=32\pi\left[\begin{array}{c}0 \\j_{0}(\alpha) \\ 0 \\ 2j_{1}(\alpha)/\alpha \\ j_{2}(\alpha)/\alpha^2\end{array}\right] ,
\end{equation}
\begin{equation}
    Y^{V(0)}=32\pi\left[\begin{array}{c}0 \\j_{0}(\alpha) \\ 0 \\j_{0}(\alpha)- j_{1}(\alpha)/\alpha \\ j_{1}(\alpha)/\alpha-4j_{2}(\alpha)/\alpha^2\end{array}\right] ,
\end{equation}
\begin{equation}
    Y^{B(0)}=32\pi\left[\begin{array}{c}j_{0}(\alpha) \\j_{0}(\alpha) \\ 2j_{1}(\alpha)/\alpha 
        \\2j_{1}(\alpha)/\alpha \\ 2j_{2}(\alpha)/\alpha^2\end{array}\right] ,
\end{equation}
\begin{equation}
    Y^{L(0)}=16\pi\left[\begin{array}{c}j_{0}(\alpha) \\2j_{0}(\alpha) \\ 2j_{1}(\alpha)/\alpha -2j_{0}(\alpha)
        \\4j_{1}(\alpha)/\alpha -4j_{0}(\alpha) \\ \left(8/\alpha^2-1\right)j_{2}(\alpha) -j_{1}(\alpha)/\alpha\end{array}\right].
\end{equation}
So the ORF of any TDI channel in the small antenna limit is 
\begin{equation}
    \begin{aligned} 
        \Gamma_{TDI}^{A0}(f)&= \frac{e^{\frac{i}{2}\left(\beta_J-\beta_I\right)}}{32 \pi} \sum_{i, j} P_{I, i} P_{J, j}
        \left\{A^{A(0)}(\alpha_{ij}) \right.\\
        &+2 B^{A(0)}(\alpha_{ij})\left(1+\hat{u}_{I,i} \cdot \hat{u}_{J,j}\right)\\
        &+C^{A(0)}(\alpha_{ij})\left(\left(\hat{u}_{I,i} \cdot \hat{s}_{ij}\right)^{2}
        +\left(\hat{u}_{J,j} \cdot \hat{s}_{ij}\right)^{2}\right)\\ &
        +4 D^{A(0)}(\alpha_{ij})\left(\hat{u}_{I,i} \cdot \hat{u}_{J,j}\right)\left(\hat{u}_{I,i} \cdot \hat{s}_{ij}\right)
        \left(\hat{u}_{J,j} \cdot \hat{s}_{ij}\right) \\
        &\left.+E^{A(0)}(\alpha_{ij})\left(\hat{u}_{I,i} \cdot \hat{s}_{ij}\right)^{2}\left(\hat{u}_{J,j} \cdot \hat{s}_{ij}\right)^{2}\right\} .
    \end{aligned} 
\end{equation}
Notice that there is an extra phase factor $e^{\frac{i}{2}\left(\beta_J-\beta_I\right)}$, due to the difference in the arm lengths of LISA and TianQin.
For LISA-TianQin network, the phase factor $e^{i2\pi f/65\text{mHz}}$ can not ne ignored in the detection band.

Since the small antenna limit may not work well, we extend it slightly.
We expand the eq.~(\ref{Gamma_abcd}) as a Taylor series of the frequency $f$.
To the zeroth order term is the expression for the small antenna approximation above.
Expanding to the next order, we obtain that
\begin{equation}\label{Gamma2}
    \begin{aligned}
    \Gamma_{a b c d}^{A2}&\left(\alpha_{ij}, \beta_I, \beta_J, \hat{u}_{I,i}, \hat{u}_{J,j}, \hat{s}_{i j}\right)
    =\Gamma_{a b c d}^{A(0)}\left(\alpha_{ij}, \hat{s}_{i j}\right)\\
    &+\Gamma_{a b c d}^{A(2)}\left(\alpha_{ij}, \beta_I, \hat{u}_{I,i}, \hat{s}_{i j}\right)
    +\Gamma_{a b c d}^{A(2)}\left(\alpha_{ij}, \beta_J, \hat{u}_{J,j}, \hat{s}_{i j}\right), 
    \end{aligned}
\end{equation}
where 
\begin{equation}
    \begin{aligned}
    \Gamma_{d b c d}^{A(2)}(\alpha, \beta, \hat{u}, \hat{s}) 
    &=-\frac{1}{6} \int d^{2} \Omega_{\hat{n}}\left(\frac{\beta}{2}[1+\hat{n} \cdot \hat{u}]\right)^{2} \\
    &\times \sum_{A} e_{a b}^{A} e_{c d}^{A} e^{-i \alpha \hat{n} \cdot \hat{s}} .
    \end{aligned}
\end{equation}
Following the same method, we construct it with $\delta_{ab}$, $s_{a}$ and $u_a$ under the premise of ensuring its symmetry.
Then solve the linear equations for the coefficients, we can get the ORF of the second order. 
The details of calculation are provided in appendix~\ref{appa}.

\begin{figure}[!t]
    \centering
    \includegraphics[width=0.45\textwidth]{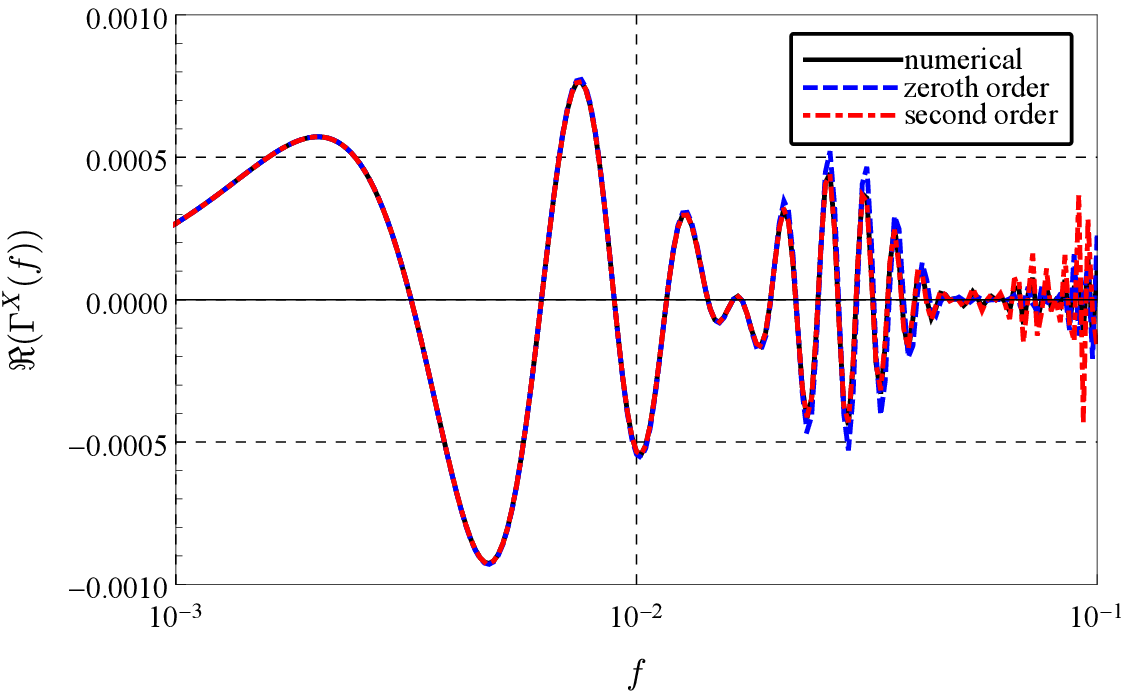} 
    \includegraphics[width=0.45\textwidth]{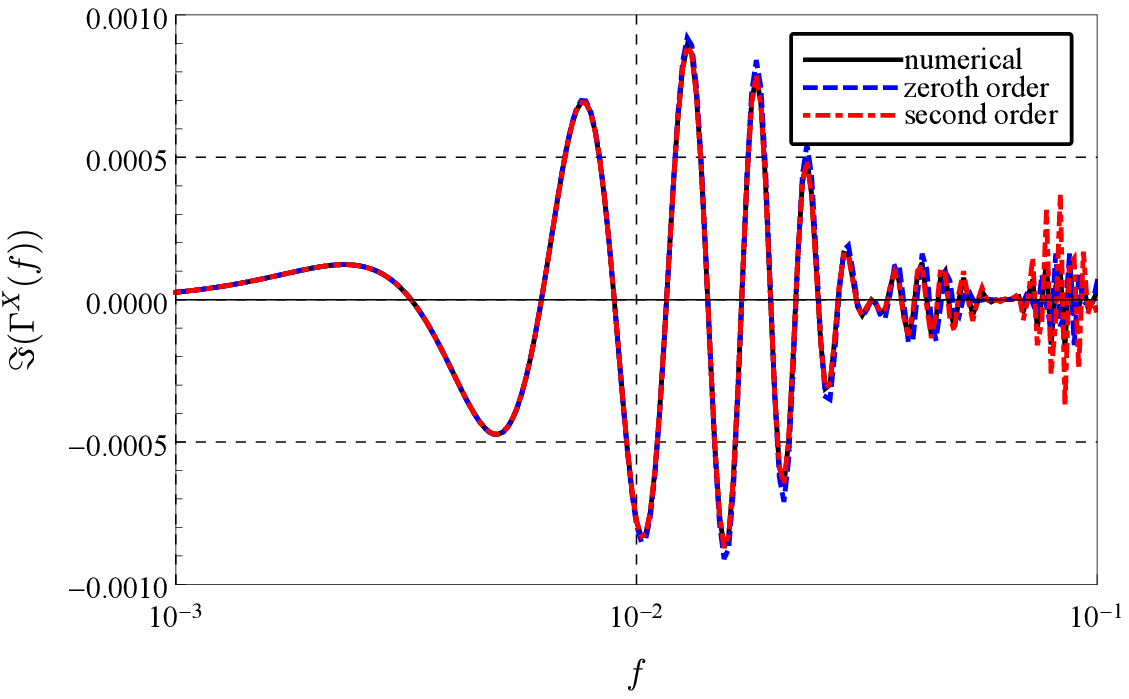}
    \caption{The tensor ORF of the $X$ channel for LISA-TianQin network. The real and imaginary parts of ORF are the left and right respectively, and the time is $t=0$. }\label{fig3}
\end{figure} 

\begin{figure*}[!t]
    \centering
    \includegraphics[width=1\textwidth]{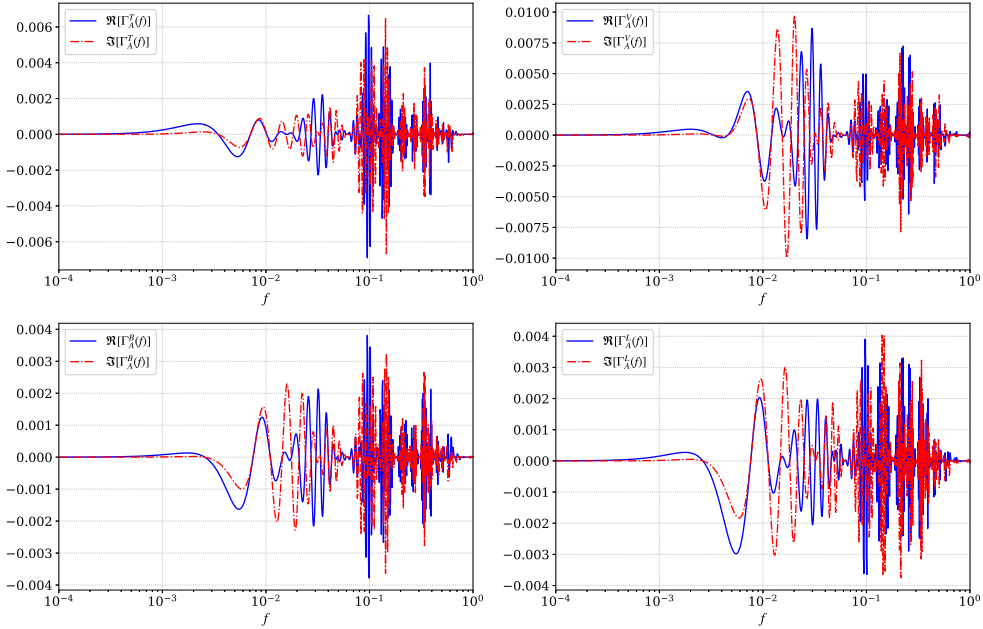}
    \caption{The different polarization ORF of the $A$ channel for LISA-TianQin network. The ORF curve is plotted for that $t=0$. 
    And the real and imaginary parts are represented by different types of curves.}\label{fig4}
\end{figure*} 

\begin{figure}[!t]
    \centering
    \includegraphics[width=1\textwidth]{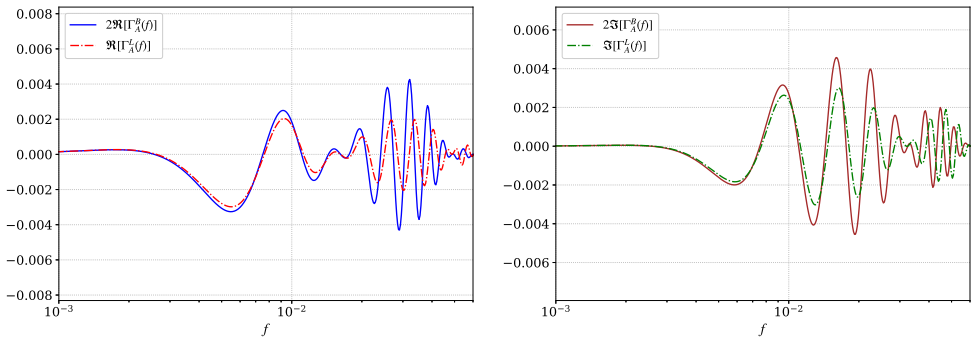}
    \caption{The comparison of the A channel ORF of two scalar modes for $t=0$.  The left-hand side is the real part, and the right-hand side is the imaginary part. }\label{fig5}
\end{figure} 

\begin{figure*}[!t]
    \centering
    \includegraphics[width=1\textwidth]{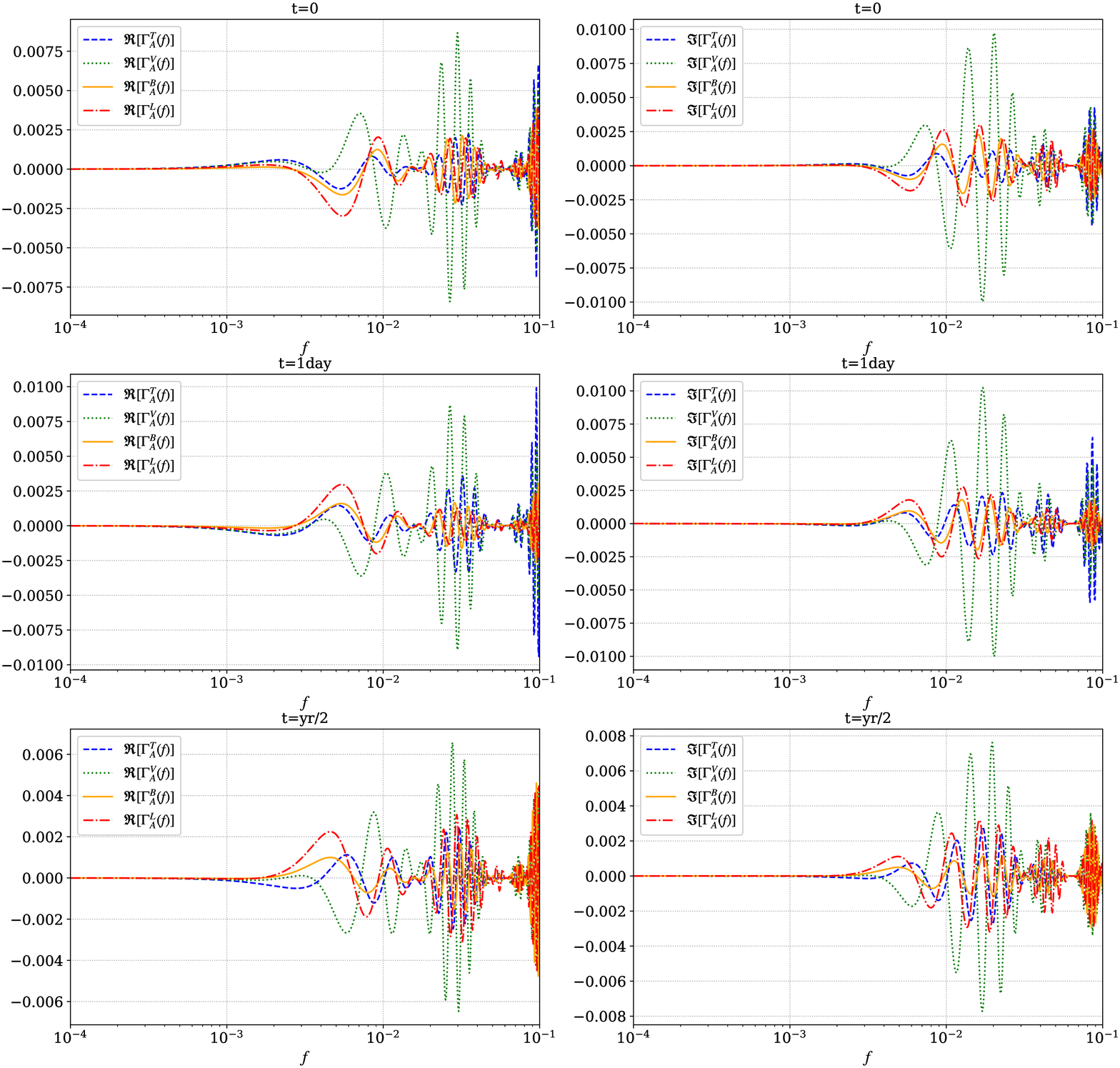}
    \caption{The ORF at different times. The left and right sides from top to bottom are the real and imaginary parts of the ORF at $t=0$, $t=1\text{day}$ and $t=\text{yr}/2$ respectively.}\label{fig6}
\end{figure*}

In order to quantify the accuracy of the expand ORF, we compared the zero-order (small antenna approximation) and second-order with numerical integration, as shown in figure~\ref{fig3}, taking the $X$ channel as an example .
For frequencies below the characteristic frequency $f<c/(2L^{LS})=0.06\text{Hz}$, 
the second-order expression is more accurate than the zero-order expression, which is in good agreement with the numerical integration.
If the phase factor $e^{\frac{i}{2}\left(\beta_J-\beta_I\right)}$ is ignored, the zero-order accuracy will be even worse.
For frequencies above the characteristic frequency $f>c/(2L^{LS})=0.06\text{Hz}$, the accuracy of second-order is worse compared with zero-order, since the error of second-order grows faster than zero-order's.
To improve accuracy, we choose to splice the two at the characteristic frequency, 
\begin{equation}
    \Gamma^A_{TDI}(f) = \begin{cases}
        \Gamma_{TDI}^{A0}(f)  & f\geq c/(2L^{LS}) \\
        \Gamma_{TDI}^{A2}(f)  & f<c/(2L^{LS}) 
    \end{cases} .
\end{equation}
They are smooth at the connection point, both equal to 0.
This concatenated expression has high enough precision for SGWB data analysis.
The detector is insensitive to frequency bands above the characteristic frequency due to loud noise and low response for SGWB.
Also, even if the numerical integration method is used, the calculation for frequencies above the characteristic frequency is very slow and the result is inaccurate, due to the rapid oscillation caused by the high frequency.

In figure~\ref{fig4}, we show the ORF of $A$ channel for different polarizations.
We choose not to normalize the ORF, because normalization is thankless for the TDI combination of the LISA-TianQin network.
Unlike ground-based detectors, TDI combinations are insensitive to low frequencies because of the  virtual
equal arm interferometric measurements.
Actually, the ORF should be proportional to $\beta_I\beta_J$.
On the other hand, it is not trivial to define them co-located for any TDI channel, since LISA and TianQin have different arm lengths.
Furthermore, the ORF of different polarizations also share some common characteristics.
Their zero point distributions are similar: $f \approx nc/(2|\Delta \vec{x}|)$, $f = nc/(2L_I)$ and $f = nc/(2L_J)$.
Due to the different arm lengths of the two detectors, TianQin is still sensitive in the frequency band 
$f > c/(2L^{LS}) \approx 60\text{mHz}$, so the ORF does not decay rapidly to 0 beyond $f = c/(2L^{LS}) \approx 60\text{mHz}$.
The peak in the frequency band of 60mHz to 860mHz is mainly contributed by the response of Tianqin, although it has virtually no contribution to the SNR due to the intense noise.
As the frequency continues to increase, the ORF will continue to decay, which is determined by the properties of the spherical Bessel function.

It is well acknowledged that the angular antenna response of the Michelson interferometer-type detector to breathing and longitudinal modes is degenerate at low frequency approximations, namely, $F^L(f,\hat{n})=\sqrt{2}F^B(f,\hat{n})$.
This results in the degeneracy of the ORFs for the two scalar polarizations in the low frequency range, with a quantitative relationship of $\Gamma^L=2\Gamma^B$.
However, as the frequency increases, the degeneracy between them is removed as shown in figure~\ref{fig5}.
This implies that it is possible to resolve two scalar modes through the LISA-TianQin network.
The key point is that the degeneracy is broken, not how different the ORFs of the two scalar modes are, as distinguishing between the two modes requires different correlation channels to provide more information.
As the quantitative relationship of $\Gamma^L=2\Gamma^B$ always remains in the low frequency range, it does not provide any information to distinguish between the two scalar modes, regardless of the number of correlated channels added.
In the high frequency range, the relationship of $\Gamma^L=2\Gamma^B$ is no longer maintained, 
which means that the ratio of $\Gamma^L$ and $\Gamma^B$ for different correlation channels may vary. 
This provides the possibility of distinguishing between breathing and longitudinal modes.

What more, we plot the ORF for different polarizations at different times in figure~\ref{fig6}.
The time-varying property makes data analysis more difficult, 
but also increases the chance of resolving different polarizations.
Correlations at different locations can be used as multi-correlation channels to distinguish between the four polarizations. 
Based on this idea, an algorithm can be developed to achieve the separation of different polarizations.
And the accuracy of a polarization identification depends on how different the rate of change in time of its ORF is from other modes'.
Details are provided in section~\ref{sec5}.

\section{Sensitivity for the background of alternative polarizations}\label{sec4}
\begin{figure*}[!t]
    \centering
    \includegraphics[width=1\textwidth]{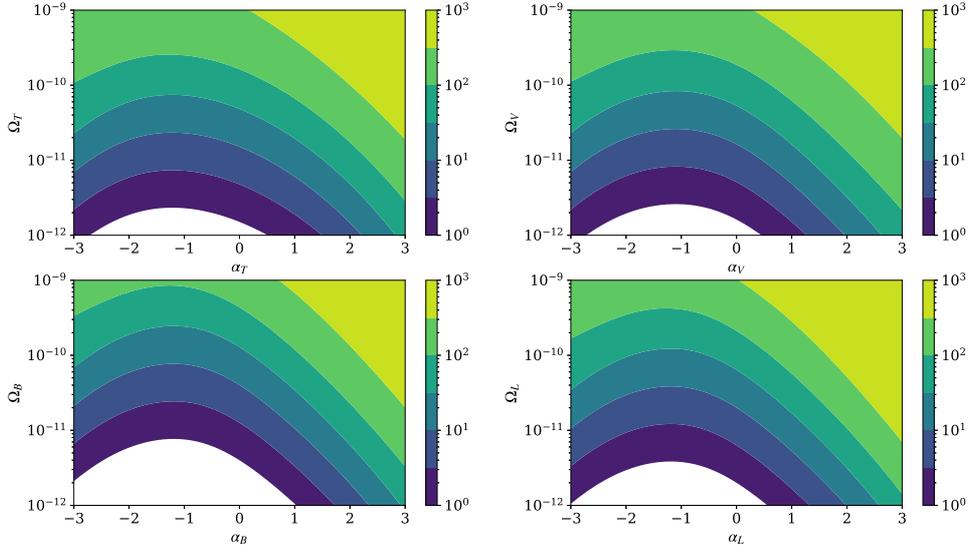}
    \caption{The SNR of power-law SGWB with different amplitude $\Omega^A$ and index $\alpha^A$ for different polarizations. The reference frequency is choose to be $f_0=1\text{mHz}$. The A channel is chosen for calculation here and the rest of the article.   }\label{fig7}
\end{figure*} 

\begin{figure}[!t]
    \centering
    \includegraphics[width=0.6\textwidth]{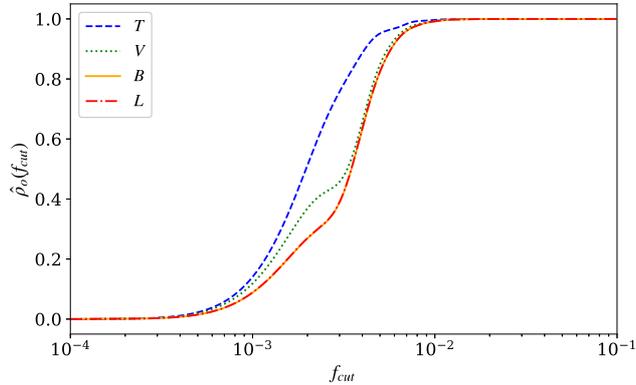}
    \caption{The normalized SNR as a function of the cutoff frequency.   }\label{fig8}
\end{figure}

\subsection{Optimal filter}
In addition to the naive method of directly correlating the output of two detecters in section~\ref{sec3},
the idea of matched filter is often used to improve the SNR.
Matched filtering starts by multiplying the correlation signal (\ref{S_f}) by a filter function,
\begin{equation}\label{Sm}
    S_m=\int^{\infty}_{-\infty}df \int^{\infty}_{-\infty}df^{\prime}\delta_T(f-f^{\prime})
    \tilde{s}_I(f) \tilde{s}^{*}_J (f^{\prime}) Q(f).
\end{equation}
The SNR is~\cite{1999prd_SGWB_detecte}
\begin{equation}
    \rho_m^2=T\left(\frac{3H_0^2}{2\pi^2}\right)^2 \frac{(Q,\frac{\sum_A \Gamma_{IJ}^A(f) \Omega^A_h(f)}{f^3P_I(f)P_J(f)})^2}{(Q,Q)} ,
\end{equation}
where the inner product of $A$ and $B$ is defined as $(A,B)\equiv \int_{-\infty}^{\infty}dfA^*(f)B(f) P_I(f)P_J(f)$.
And the optimal filter is 
\begin{equation}\label{filter}
    Q(f) \propto \frac{\sum_A \Gamma_{IJ}^A(f)\Omega^A_h(f)}{f^3P_I(f)P_J(f)},
\end{equation}
when mean that the model and the actual energy-density spectrum match.
The resulting optimal SNR is given by 
\begin{equation}\label{snr_o}
    \rho_{o} \approx \sqrt{T}\frac{3H_0^2}{2\pi^2}\left[\int^{\infty}_{-\infty}df \frac{\left|\sum_A \Gamma_{IJ}^A(f)\Omega^A_h(f) \right|^2}{f^6P_I(f)P_J(f)}\right]^{1/2} .
\end{equation}
The premise of the above expression is that the noise amplitude is much larger than the signal.
In general, it will be modified to~\cite{2001prd_SGWB_LISA}
\begin{equation}
    \rho_{o}=\sqrt{T}\frac{3H_0^2}{2\pi^2}\left[\int^{\infty}_{-\infty}df \frac{\left|\sum_A \Gamma_{IJ}^A(f)\Omega^A_h(f) \right|^2}{f^6 M(f)}\right]^{1/2} ,
\end{equation}
where
\begin{equation}
\begin{aligned}
    M(f) &= P_I(f)P_J(f) + S_h(f)\left(P_I(f)\Gamma_{JJ}(f) + P_J(f)\Gamma_{II}(f)\right) \\
    &+ S^2_h(f)\left(\Gamma^2_{IJ}(f)+\Gamma_{II}(f)\Gamma_{JJ}(f)\right).
\end{aligned}
\end{equation}

Once again, considering the time-varying effect of LISA-TianQin, 
the total SNR after segmented processing is
\begin{equation}
    \rho_{o,total}=\sqrt{T}\frac{3H_0^2}{2\pi^2}\left[\int^{\infty}_{-\infty}df \frac{\left|\sum_A \tilde{\Gamma}_{IJ}^A(f)\Omega^A_h(f) \right|^2}{f^6 M(f)}\right]^{1/2} ,
\end{equation}
where the average ORFs is defined as 
\begin{equation}
    \tilde{\Gamma}_{IJ}^A(f) = \frac{1}{N}\sqrt{\sum_{k=1}^N |\Gamma_{IJ,k}^A(f)|^2 }.
\end{equation}

Assuming that the spectrum of SGWB is power-law, the optimal SNR calculated by eq.~(\ref{snr_o}) can be used to evaluate the detection capability of LISA-TianQin network for SGWB.
We show the SNR of SGWB with different polarizations in figure~\ref{fig7}, 
where the observation time is chosen as $T=1\text{yr}$ and the TDI channel is selected as channel A here and the rest of the article.
The result shows that the LISA-TianQin network is more sensitive to tensor and vector polarizations than the two scalar polarizations.
For example, for the same spectral index $\alpha^A=2/3$, in order to achieve a SNR of 10, the amplitudes at reference frequency for different polarizations are required to be: $\Omega_0^T=8.59 \times 10^{-12}$, $\Omega_0^V=7.58 \times 10^{-12}$, 
$\Omega_0^B=1.71 \times 10^{-11}$ and $\Omega_0^L=8.61 \times 10^{-12}$ .
In addition, in order to find the most sensitive frequency band of LISA-TianQin, we define a normalized SNR as a function of the cutoff frequency,
\begin{equation}
    \hat{\rho}_{o}(f_{\text{cut}})=\frac{1}{\rho_{o}}\sqrt{T}\frac{3H_0^2}{2\pi^2}\left[2\int^{f_{\text{cut}}}_{0}df \frac{\left|\sum_A \Gamma^A(f)\Omega^A_h(f) \right|^2}{f^6P_I(f)P_J(f)}\right]^{1/2} .
\end{equation}
We show the function $\hat{\rho}_{o}(f_{\text{cut}})$ in figure~\ref{fig8} where the power-law SGWB with spectrum index $\alpha_A=2/3$ are assumed.
The results show that the frequency band that contributes the most to the SNR of background from compact binary coalescences is $1\text{mHz}-10\text{mHz}$.
The contribution of high-frequency bands to the SNR is negligible, 
with each polarization SNR of more than 0.06Hz accounting for $8.5\times10^{-8}$, $6.5\times10^{-8}$, $7.7\times10^{-8}$ 
and $1.1\times10^{-7}$, respectively. 
This reflects the fact that strong noise prevents high-frequency bands from being accurately detected at present.
The error of the ORF in high frequency region does not matter.
Of course, as detector technology advances, noise levels decrease,
the development of high-precision ORFs in high frequency region is necessary to detect potential high-frequency sources.
Also note that the curves for the scalar-breathing and the scalar-longitudinal mode are the same, since their responses are the same at low frequency approximations.
If we increase the spectrum index, which means that the high frequency part of the SGWB is louder, the curves for the scalar-breathing and the scalar-longitudinal mode will be differentiated.

\subsection{Parameter estimation accuracy}
\begin{figure}[!t]
    \centering
    \includegraphics[width=1\textwidth]{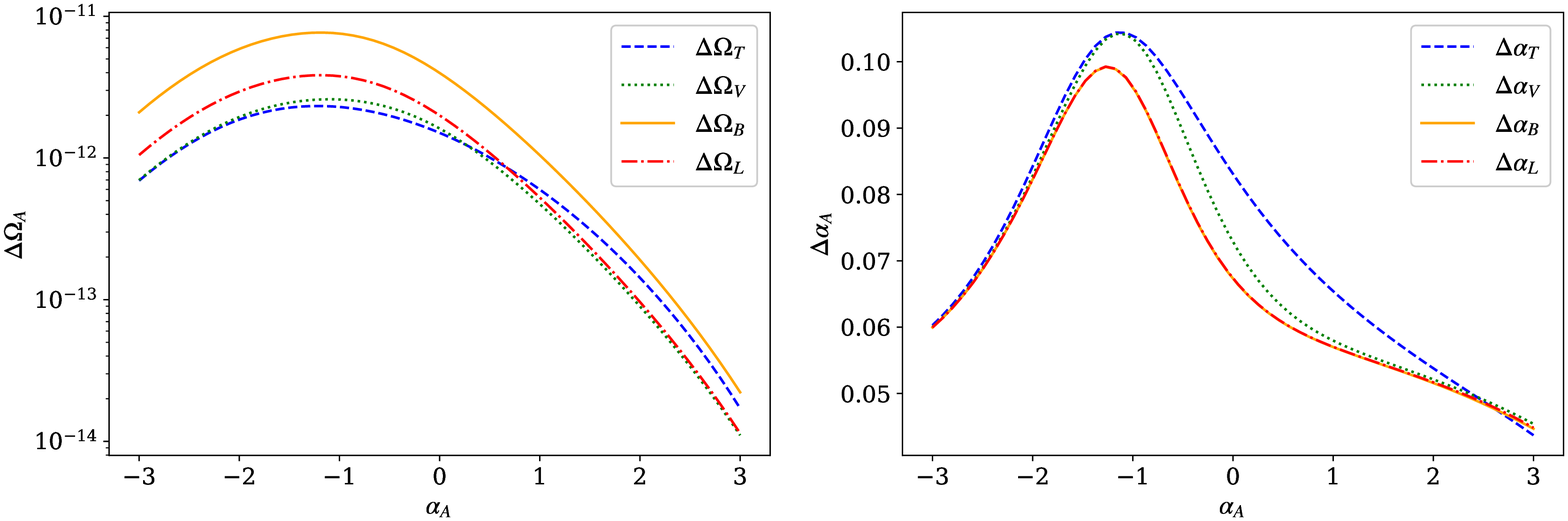}
    \caption{The uncertainties of the amplitude $\Omega^A$ and index $\alpha^A$ for different polarizations. The uncertainties of the index $\alpha^A$ are calculated with suitable amplitude parameters such that the SNR is 10.  }\label{fig9}
\end{figure} 
Under the premise that the model is accurate, the optimal filter can not only improve the signal-to-noise ratio, but also provide an estimate of the model parameters.
When the SNR is high enough, the Fisher information matrix (FIM) can be used to evaluate parameter estimation accuracy~\cite{2007prd_LISA_parameter_error}.
The FIM is defined by 
\begin{equation}
    F_{ab}=T \int^{\infty}_{-\infty}df 
    \frac{(\sum_A \Gamma^A(f)\frac{\partial S^A_h(f)}{\partial \theta_a})(\sum_A \Gamma^{A*}(f)\frac{\partial S^A_h(f)}{\partial \theta_b})}{P_I(f)P_J(f)} ,
\end{equation}
and the estimate error of a parameter $\Delta \theta_a$ is estimated from the inverse of FIM 
\begin{equation}
    \Delta \theta_a =\sqrt{F_{aa}^{-1}}.
\end{equation}
There are a total of eight parameters we need to estimate for a power law model that includes all polarizations.
And the parameter estimation errors are given by 
\begin{equation}
    \begin{aligned}
    \Delta\Omega^A_0 &= \frac{1}{\sqrt{T}} \frac{2\pi^2}{3H_0^2} \left[\int^{\infty}_{-\infty}df 
    \frac{ |\Gamma^A(f)|^2(f/f_0)^{2\alpha_A}}{f^6P_I(f)P_J(f)}\right]^{-1/2}, \\
    \Delta\alpha_A &= \frac{1}{\sqrt{T}} \frac{2\pi^2}{3H_0^2} \frac{1}{\Omega^A_0} 
     \left[2\int^{\infty}_{-\infty}df 
    \frac{ |\Gamma^A(f)|^2(f/f_0)^{2\alpha_A}\ln(f/f_0)}{f^6P_I(f)P_J(f)}\right]^{-1/2}.
    \end{aligned}
\end{equation}

The uncertainties of the amplitude $\Omega^A$ and index $\alpha^A$ for different polarizations are shown in figure~\ref{fig9}.
In general, the accuracies of the amplitude parameters for tensor and vector are higher than the two scalar polarizations.
This means they are easier to detect.
The uncertainties of the index are plotted with suitable amplitude parameters such that the SNR is 10.
In such a case, the spectral parameter estimation accuracy of the two scalar modes is higher.
Another point worth noting is that the uncertainty curves of the scalar-breathing mode and the scalar-longitudinal mode almost overlap because their ORFs are degenerate at low frequencies.
And when the exponent increases, which means that the high frequency signal is stronger, 
the curves are not overlapping, because the degeneracy breaks at high frequency.

\section{Polarization separation}\label{sec5}
If the alternative polarization really exists, 
different polarization modes will be mixed in the cross-correlation signal.
We need suitable methods to distinguish difference modes in the cross-correlation signal.
One possible approach is Bayesian model selection proposed in~\cite{2017prx_polarization_SGWB}, 
which is adopted to detect alternative polarization background in the cross-correlation data of ground-based detecter~\cite{2018_nonGR_GWB_LIGO_O1, 2019prd_LIGO_O2_GWB, 2021_GWB_O3}.
This approach is efficient and easy to implement.  
However, it may be prone to bias if the model does not fit well with the true background.

If one want to get away from being bound by model assumptions, 
there is a intuitive method to separate polarizations using multiple detecters pairs~\cite{2009prd_nonGR_GWB_LIGO}.
Different detector pairs have different ORFs, which breaks the degeneracy between different polarizations and provides enough degrees of freedom to separate different polarizations.
Notably, this approach can be implemented independently for each frequency bin.
The accuracy of the results for each frequency bin depends on the corresponding signal and noise strength. 
However, this method requires a higher signal-to-noise than the previous method.
Another disadvantage is that multiple detector pairs naturally mean that more than two detectors are required.

Since the LISA-TianQin network varies with orbits,
their ORFs for difference polarizations varies accordingly as see in section~\ref{sec3}.
So their correlation signals at different positions respond differently to different polarizations,
which means that it can be equivalently regarded as different detector pairs at different times.
There is an opportunity to extract different polarization patterns from the data from the two detectors,
using a similar approach to the multi-detector based approach~\cite{2009prd_nonGR_GWB_LIGO}.

The key point is to break the coupling of different polarizations in the data through multiple correlation channels.
If data for $n$ detectors is available, a total of $N=n(n-1)/2$ correlation channels can be constructed as 
\begin{equation}
    Z_{ij}(f) \propto \tilde{s}_{i}(f) \tilde{s}^{*}_{j}(f),
\end{equation}
where $1 \leq i < j \leq n$.
These statistics satisfy 
\begin{equation}\label{multi_correlation}
    \left\langle \bm{Z(f)} \right\rangle=\bm{M}
    \begin{bmatrix}
        \Omega^T(f) \\ \Omega^V(f) \\ \Omega^B(f) \\ \Omega^L(f)
    \end{bmatrix},
\end{equation}
where $\bm{Z(f)}=[Z_{12}(f), Z_{13}(f), ... , Z_{(n-1)n}(f)]^T$ and 
\begin{equation}
    \bm{M}=
    \begin{bmatrix}
        \Gamma^T_{12} & \Gamma^V_{12} & \Gamma^B_{12} & \Gamma^L_{12} \\
        \Gamma^T_{13} & \Gamma^V_{13} & \Gamma^B_{13} & \Gamma^L_{13} \\
        ... & ... & ... & ... \\
        \Gamma^T_{(n-1)n} & \Gamma^V_{(n-1)n} & \Gamma^B_{(n-1)n} & \Gamma^L_{(n-1)n} \\
    \end{bmatrix} 
\end{equation}
is the detecter correlation matrix.
If the rank of the matrix is greater than or equal to 4, $\text{rank}(\bm{M})\geq 4$, 
the polarization components can be separated by inverting eq.~(\ref{multi_correlation}), namely,
\begin{equation}\label{invet}
    \begin{bmatrix}
        \hat{\Omega}^T(f) \\ \hat{\Omega}^V(f) \\ \hat{\Omega}^B(f) \\ \hat{\Omega}^L(f)
    \end{bmatrix}
    = \bm{M(f)}^{-1} \left\langle \bm{Z(f)} \right\rangle .
\end{equation}
In this way, the spectrum of each polarization component can be reconstructed from the detector data by eq.~(\ref{invet}).
However, the degeneracy of ORFs of two scalar polarizations for ground-based causes the rank of the detector correlation matrix to be at most 3.
So the SGWB detected by the ground-based detector network can only be separated into tensor, vector and scalar components.

Two space-based detectors may achieve the capabilities of the above-mentioned ground-based detector network to distinguish polarizations in SGWB.
It is only necessary to replace the different correlation channels with the correlation of the two space-based detectors at different positions.
Below we use another description of maximum likelihood estimation mentioned in~\cite{2017lrr_detection_GWB} to obtain the same result.
First, the data is divided into $N$ segments (indexed by $i$), and the duration of each segment of data $\Delta T$ satisfies the condition that it is greater than the light travel time between the two detectors and less than the time scale of the ORF changes.
A cross-correlation statistic can be constructed as
\begin{equation}\label{C_i}
    \hat{C}_{i}(f)=\frac{2}{\Delta T} \frac{2 \pi^{2}}{3 H_{0}^{2}} f^{3} \tilde{s}_{1, i}(f) \tilde{s}^{*}_{2, i}(f) ,
\end{equation}
normalized such that the statistic's mean is
\begin{equation}
    \left\langle\hat{C}_{i}(f)\right\rangle=\sum_{A} \Gamma_i^{A}(f) \Omega^{A}(f) ,
\end{equation}
where $\Gamma_i^{A}(f)$ is the ORF for i-th segment data.
The statistic $\bm{\hat{C}(f)}= [\hat{C}_{1}(f), \hat{C}_{2}(f), ..., \hat{C}_{N}(f)]^T$ acts as the role of different correlation channels $\bm{Z(f)}$.
And the variance is 
\begin{equation}\label{var}
    \sigma_{i}^{2}(f)=\frac{1}{2 \Delta T \Delta f}\left(\frac{2 \pi^{2}}{3 H_{0}^{2}}\right)^{2} f^{6} P_{1, i}(f) P_{2, i}(f) 
\end{equation}
where $\Delta f$ is the frequency bin width, $P_{1, i}(f)$ and $P_{2, i}(f)$ are the noise power spectral density of detectors.

The likelihood function for $\bm{\hat{C}(f)}$ is
\begin{equation}
    \mathcal{L}[\bm{\hat{C}(f)}|\mathcal{A}] \propto \text{exp}\left[-\sum_i^N\sum_f \frac{|\hat{C}_i(f)-\sum_{A} \Gamma_i^{A}(f) \Omega^{A}(f)|^2}{2\sigma_{i}^{2}(f)} \right],
\end{equation}
where
$\mathcal{A}=[\Omega^T(f) , \Omega^V(f) , \Omega^B(f) , \Omega^L(f) ]^T$
represent the fractional energy density of different polarizations.
Equivalently, we can write it as
\begin{equation}
    \mathcal{L}[\bm{\hat{C}(f)}|\mathcal{A}] \propto \exp \left[-\frac{1}{2}(\bm{\hat{C}}- \bm{M} \mathcal{A})^{\dagger} \mathcal{N}^{-1}(\bm{\hat{C}}- \bm{M} \mathcal{A})\right] ,
\end{equation}
where the detector correlation matrix becomes
\begin{equation}\label{M}
    \bm{M}= 
    \begin{bmatrix}
        \Gamma^T_1 & \Gamma^V_1 & \Gamma^B_1 & \Gamma^L_1 \\
        \Gamma^T_2 & \Gamma^V_2 & \Gamma^B_2 & \Gamma^L_2 \\
        ... & ... & ... & ... \\
        \Gamma^T_N & \Gamma^V_N & \Gamma^B_N & \Gamma^L_N \\
    \end{bmatrix} .
\end{equation}
And the noise covariance matrix is $\mathcal{N}_{i j} = \delta_{i j} \sigma_{i}^{2}(f)$.
\begin{figure}[!t]
    \centering
    \includegraphics[width=0.6\textwidth]{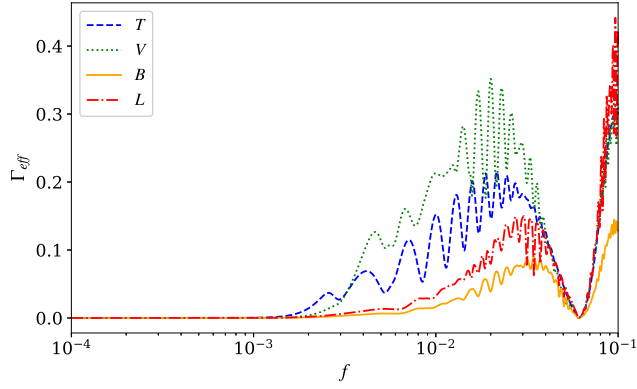}
    \caption{The efficient overlap functions for different polarizations.  }\label{fig10}
\end{figure} 

\begin{figure}[!t]
    \centering
    \includegraphics[width=0.6\textwidth]{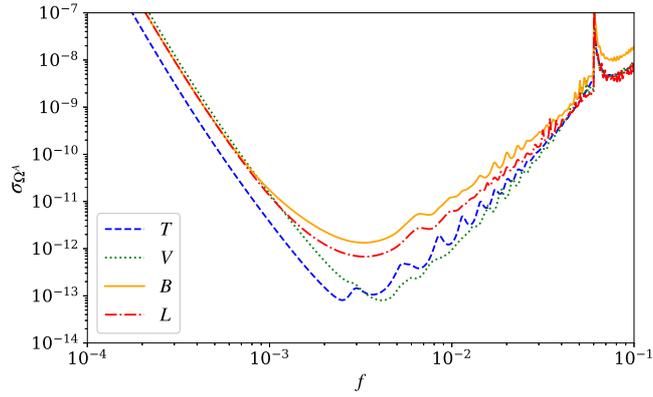}
    \caption{Equivalent sensitivity curves of different polarization components.  }\label{fig11}
\end{figure} 

\begin{figure*}[!t]
    \centering
    \includegraphics[width=1\textwidth]{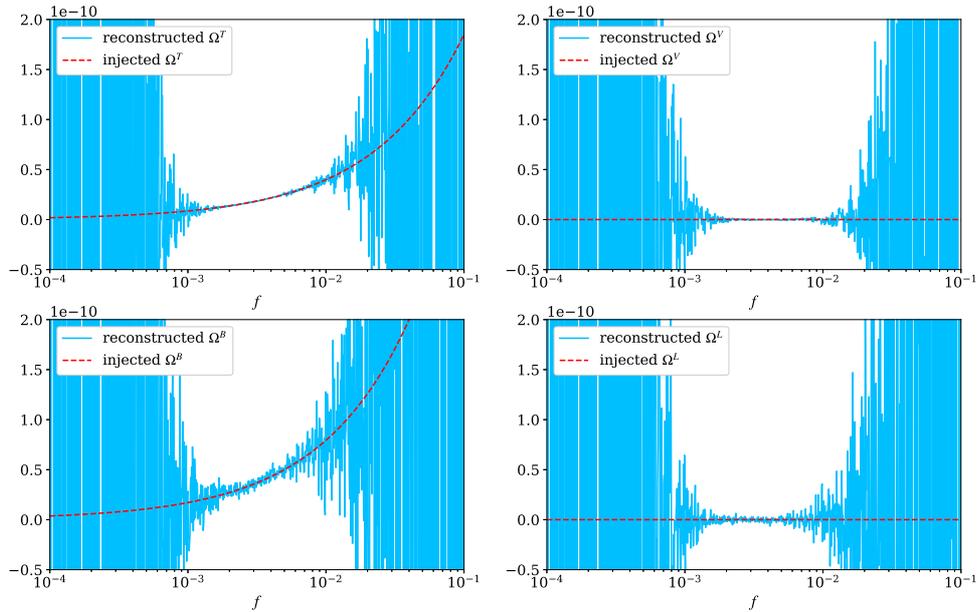}
    \caption{The reconstructed spectrum for each polarizations by the equivalent multi-detector-based method.
    The data contains tensor and breathing polarisations: $\Omega^T=8.59 \times 10^{-12}(f/1\text{mHz})^{2/3}$ and $\Omega^B=1.71 \times 10^{-11}(f/1\text{mHz})^{2/3}$.
    The red dashed line represents the actual injected spectrum,
    and the light blue solid line represents the reconstructed spectrum.}\label{fig12}
\end{figure*} 

The maximum-likelihood estimator for the fractional energy density of different polarizations are~\cite{2017lrr_detection_GWB}
\begin{equation}\label{MLE}
    \hat{\mathcal{A}}=F^{-1}X ,
\end{equation}
where $F=M^{\dagger}\mathcal{N}^{-1}M$, $X=M^{\dagger}\mathcal{N}^{-1}\hat{C}$ are the Fisher matrices and 'dirty' map for this analysis.
The maximum-likelihood estimator $\hat{\mathcal{A}}$ is equivalent to eq.~(\ref{invet}), 
which is easier to calculate in practice and avoids finding the inverse of a huge matrix.
After the data is available, the reconstructed spectrum of each polarization component can be obtained by calculating eq.~(\ref{MLE}) for each frequency.
The error will be determined by the detector noise at that frequency and the degeneracy of ORFs of different polarizations.
Specifically, the marginalized uncertainties of the maximum-likelihood estimates are given by the inverse of the Fisher matrices,
\begin{equation}
    \sigma^2_{\Omega^A} = \left\langle \mathcal{A} \mathcal{A}^{\dagger} \right\rangle _{AA}  = (F^{-1})_{AA}. 
\end{equation}
Without loss of general, we assume that the noise power spectrum for different time segments are equal.
The Fisher matrix is 
\begin{equation}
    F_{AB} = \sum_i \Gamma_i^{A} \Gamma_i^{B*}/\sigma_i^2.
\end{equation}
And the efficient overlap function can be defined by 
\begin{equation}
    \Gamma_{\text{eff}}^{A}(f) \equiv \sigma^{-1}_{\Omega^A} \sigma_i = \frac{\sigma_i}{\sqrt{(F^{-1})_{AA}}},
\end{equation}
which can reflect the intensity of the response to different polarization components.
The efficient overlap functions for LISA-TianQin network are shown in figure~\ref{fig10}.
The effective overlap functions of the two scalar components are significantly lower than that of tensors and vectors, 
due to the similar ORFs of scalar-breathing and scalar-longitudinal modes.
In order to quantify the sensitivity of this method to different polarization components, detector noise should also be included.
This is the role of uncertainties $\sigma_{\Omega^A}$, which are shown in figure~\ref{fig11}.
For a certain frequency $f$, the intensity of $A$ component must meet $\Omega^A(f)>\sigma_{\Omega^A}(f)$ to be recognized.

To visually measure the method's ability to distinguish between different polarizations, 
an overall SNR can be defined as 
\begin{equation}
    \begin{aligned}
        \text{SNR}_{\text{eff}}^A &= \left[\int_0^\infty df \frac{|\Gamma^A_{\text{eff}}(f)\Omega_h^A(f)|^2}{\sigma_i^2}\right]^{1/2} \\
        &=\sqrt{\Delta T}\frac{3H_0^2}{2\pi^2}\left[2\int^{\infty}_{0}df \frac{\left|\Gamma_{\text{eff}}^A(f)\Omega^A_h(f) \right|^2}{f^6P_I(f)P_J(f)}\right]^{1/2} .
    \end{aligned}
\end{equation}
By convention, a SNR of 10 is considered to be a signal extract from the data.
For an astronomical background spectrum with index $\alpha=2/3$, the amplitudes at reference frequency, $f_0=1\text{mHz}$,
for different polarizations are required to be: 
$\Omega_0^T=1.27 \times 10^{-11}$, $\Omega_0^V=9.55 \times 10^{-12}$, 
$\Omega_0^B=1.43 \times 10^{-10}$ and $\Omega_0^L=7.23 \times 10^{-11}$ .
Comparing with the results of optimal SNR, 
we find that the ability of the method to resolve vector polarization is the best, followed by tensor. 
The ability to distinguish two scalars is the worst, and the SNR is equivalent to 12\% of the optimal SNR.
For comparison, for the flat spectral index $\alpha=0$, the amplitudes for different polarizations need to exceed:
$\Omega_0^T=2.90 \times 10^{-11}$, $\Omega_0^V=2.59 \times 10^{-11}$, 
$\Omega_0^B=3.61 \times 10^{-10}$ and $\Omega_0^L=1.84 \times 10^{-10}$ .

To verify the ability of the above method in resolving polarization modes, 
we apply it to the simulated data of different situations.
Notably, we simulated a mixture of tensor and breathing modes to examine whether it can distinguish between breathing and longitudinal modes.
We simulate a year of data and divided it into 3600 segments, 
and the noise is generated according to eq.~(\ref{var}).
The spectrum injected into each piece of data are 
$\Omega^T=8.59 \times 10^{-12}(f/1\text{mHz})^{2/3}$ and $\Omega^B=1.71 \times 10^{-11}(f/1\text{mHz})^{2/3}$.
The parameters are choose such that individual tensor and scalar-breathing components both correspond to an optimal SNR of 10, 
mixed together with an optimal SNR of 15.
The spectrum for each polarization reconstructed with eq.~(\ref{MLE}) is shown in figure~\ref{fig12}.
The spectral accuracy of the reconstructed spectrum in the frequency domain $10^{-3}\text{Hz}-10^{-2}\text{Hz}$ is relatively high.
It can be concluded that in the most sensitive frequency band, the different modes can indeed be distinguished.
Even two scalar modes are capable of distinguishing, but the accuracy is worse than tensor or vector modes.

\begin{figure*}[!t]
    \centering
    \includegraphics[width=1\textwidth]{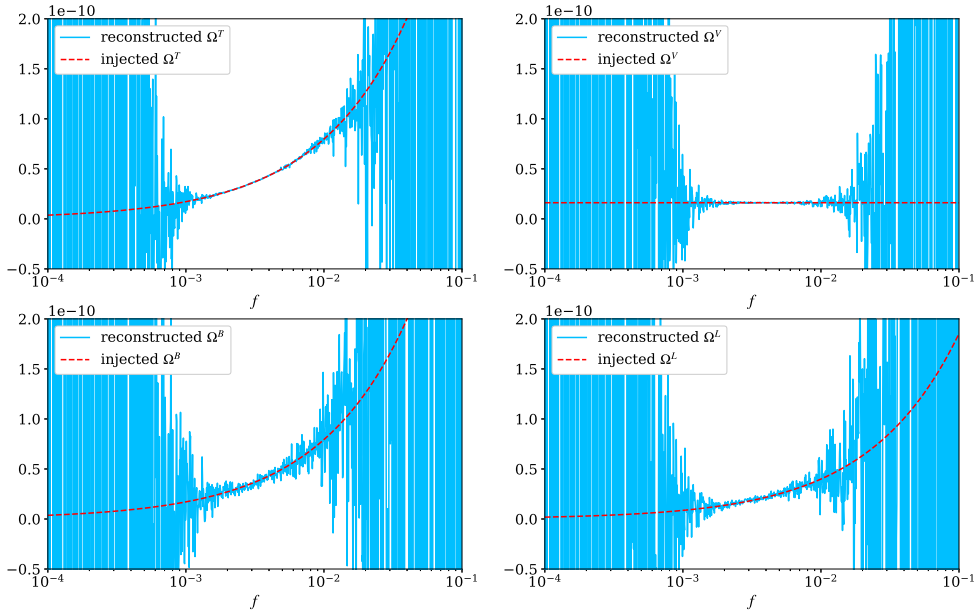}
    \caption{The reconstructed spectrum for each polarizations by the equivalent multi-detector-based method in the case of all four modes.
    The spectra of each polarization are: $\Omega^T=1.42 \times 10^{-11}(f/1\text{mHz})^{2/3}$, $\Omega^V=1.62 \times 10^{-11}$,
    $\Omega^B=1.71 \times 10^{-11}(f/1\text{mHz})^{2/3}$ and $\Omega^L=8.61 \times 10^{-12}(f/1\text{mHz})^{2/3}$.
    }\label{fig13}
\end{figure*} 

\begin{figure*}[!t]
    \centering
    \includegraphics[width=1\textwidth]{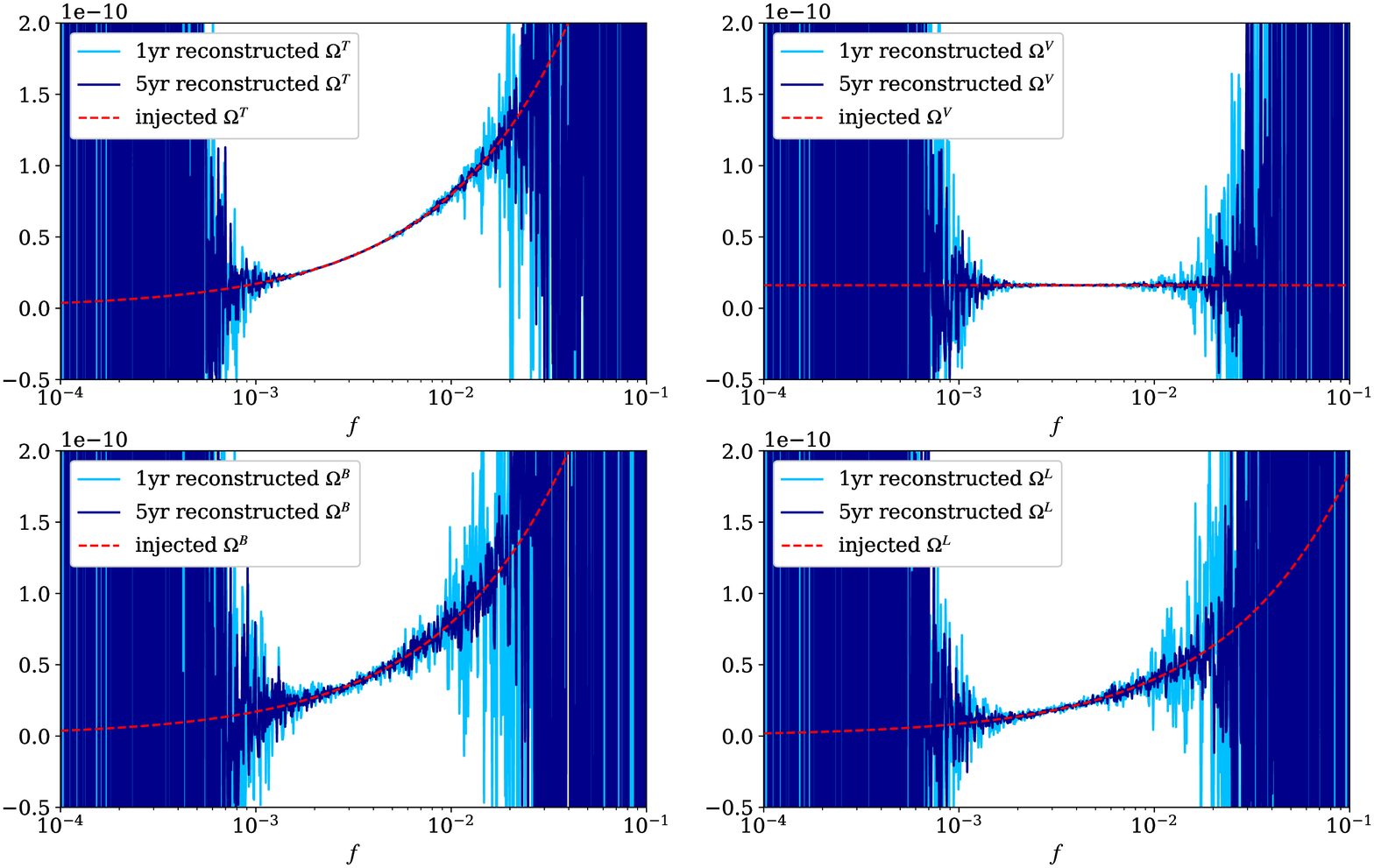}
    \caption{The comparison of the reconstructed spectrum with observation time $T=1 \text{years}$ and $T=5 \text{years}$.
    The injected spectra are the same as in figure~\ref{fig13}.
    }\label{fig14}
\end{figure*} 

As discussed above, the separate breathing mode of the two scalar modes will not be mistaken for longitudinal mode.
Furthermore, some alternative theories of gravity would predict more complex mixed polarizations.
On the other hand, the spectral index for different polarizations may be different.
Therefore, we simulate a more complex case with all four polarizations at the same time, in which the spectral index of the vector mode is set to distinguish it from the other three.
The special spectrum injected into the data are:
$\Omega^T=1.42 \times 10^{-11}(f/1\text{mHz})^{2/3}$, $\Omega^V=1.62 \times 10^{-11}$,
$\Omega^B=1.71 \times 10^{-11}(f/1\text{mHz})^{2/3}$ and $\Omega^L=8.61 \times 10^{-12}(f/1\text{mHz})^{2/3}$.
The indices for tensor, breathing and longitudinal modes are set to $2/3$, 
while the spectra of vector mode is set to be flat, namely, $\alpha_V=0$. 
As shown in figure~\ref{fig13}, the spectra of each polarization are correctly reconstructed.
The noise of the breathing and longitudinal modes is significantly greater than that of the other two modes,
but each can be identified independently.
The reconstructed polarization spectrum fall exactly near the injected true value, and the difference between the polarizations is only the strength of noise.

figure~\ref{fig14} shows the result of increasing the observation time for the same signal, where the parameters of the signal are exactly the same as figure~\ref{fig13}, and the observation time is increased to 5 years.
The SNR increases with observation time, specifically $\text{SNR} \propto \sqrt{T}$.
Equivalently, five times the observation time would reduce the uncertainty by a factor of 2.2.
This behavior is clearly presented in figure~\ref{fig14}.
The noise of the reconstruction spectrum observed over five years is significantly reduced for the same signal.

The simulation results verify the ability of the equivalent multi-detector-based method to separate different polarizations.
The reconstructed spectrum for each polarization obtain by this method falls near the true value, 
and the uncertainty depends on the noise of the detector and the correlation between one component and other polarization component in the correlation matrix in eq.~(\ref{M}).
It may be misunderstood that this correlation between different polarized components refers to their ORF similarity.
In fact, it does not matter how many times the ORF for one mode is that for the other, 
what matters is how different the values of each of them change at different positions in orbit.
For example, the degeneracy of breathing and longitudinal modes in the low frequency band results in the ORF  for longitudinal mode being twice that of the breathing mode.
But the key point is that this double relationship is always maintained in the low frequency band, 
regardless of how the detectors move.
For high frequency bands, the degenerate relation is broken, and the ORFs of breathing and longitudinal modes have the opportunity to vary with time differently.
The time-varying effect has been encapsulated in the equivalent ORFs $\Gamma_{\text{eff}}^{A}(f)$ constructed by the inverse of correlation matrix, which could quantifies the ability to separate polarization.

\section{Discussion}\label{sec6}
In this paper, we studied the detectability of alternative polarizations of SGWB with LISA-TianQin network.
The different orbits of LISA and TianQin make them very different from ground-based detectors.
The relative orientation of their orbital planes changes with time, which means that the ORF will vary accordingly.
In other words, the SGWB signal in their cross-correlation signal varies over time.
This will pose certain challenges for data analysis.
On the other hand, LISA-TianQin has advantages for resolving polarization modes in SGWB in principle.

TDI technology is applied to space gravitational wave detectors to suppress laser phase noise.
Based on the small antenna approximation, we obtain the ORF of the second-order expansion of the frequency for any TDI channel.
This method can be applied to any laser interferometric detector, even with different configurations and different arm lengths.
For the LISA-TianQin network, the accuracy can be effectively improved in its most sensitive frequency band.
Then we study the detectability of alternative polarizations in SGWB with LISA-TianQin network.
We calculated the signal-to-noise ratio and parameter estimation accuracy of power-law SGWB with different polarization modes.

Once the SGWB is detected, it is necessary to distinguish different polarization modes from it.
An equivalent multi-detector-based approach can be applied on the LISA-TianQin network, thanks to its special orbital variation.
The ORFs of the LISA-TianQin network are different at different times, so they can be equivalently regarded as different detectors.
In fact, the cross-correlated signals at different times form a system of linear equations for different polarization modes.
In turn, the system of equations can be solved to obtain different polarization modes, including the influence of noise of course.
Although this method does not rely on model assumptions, it requires a high signal-to-noise ratio.
Its resolution at a certain frequency bar is affected by the degeneracy of the ORFs for different polarizations in addition to noise at that frequency bar.
what's more, it has more advantages in distinguishing scalar-breathing mode and scalar-longitudinal mode.

Our proposed method for resolving polarization relies on the change of relative orientation of the two detectors over time, 
with LISA-TianQin being the most pronounced due to differences in orbital configuration.
However, our analysis can be extended to include any space-based detector.
We will compare the resolution capabilities of different detector pairs and evaluate the ability improvement of multi-detector network cooperation as future work.

\appendix
\section{Second order overlap reduction function of any TDI channel}\label{appa}
The ORF frequency second-order expansion of any TDI channel requires the calculation of coefficients
\begin{equation}\label{gamma2}
    \Gamma_{a b c d}^{A(2)}(\alpha, \beta, \hat{u}, \hat{s})=-\frac{1}{6} \int d^{2} \Omega_{\hat{n}}\left(\frac{\beta}{2}[1+\hat{n} \cdot \hat{u}]\right)^{2} \sum_{A} e_{a b}^{A} e_{c d}^{A} e^{-i \alpha \hat{n} \cdot \hat{s}} .
\end{equation}
According to the symmetry of the index, we construct it with $\delta_{ab}$, $s_{a}$ and $u_a$:
\begin{equation}\label{gamma2_con}
    \begin{aligned} 
        \Gamma_{a b c d}^{A(2)}&(\alpha, \beta, \hat{u}) =A^{A(2)} \delta_{a b} \delta_{c d}
        +B^{A(2)}\left(\delta_{a c} \delta_{b d}+\delta_{b c} \delta_{a d}\right)
        +C_{1}^{A(2)}\left(\delta_{a b} s_{c} s_{d}+\delta_{c d} s_{a} s_{b}\right) \\ 
        &+C_{2}^{A(2)}\left(\delta_{a b} u_{c} u_{d}+\delta_{c d} u_{a} u_{b}\right)
        +D_{1}^{A(2)}\left(\delta_{a c} s_{b} s_{d}+\delta_{a d} s_{b} s_{c}+\delta_{b c} s_{a} s_{d}+\delta_{b d} s_{a} s_{c}\right) \\ 
        &+D_{2}^{A(2)}\left(\delta_{a c} u_{b} u_{d}+\delta_{a d} u_{b} u_{c}+\delta_{b c} u_{a} u_{d}+\delta_{b d} u_{a} u_{c}\right)
        +E_{1}^{A(2)} s_{a} s_{b} s_{c} s_{d}
        +E_{2}^{A(2)} u_{a} u_{b} u_{c} u_{d} \\ 
        &+E_{3}^{A(2)}\left(u_{a} u_{b} s_{c} s_{d}+s_{a} s_{b} u_{c} u_{d}\right)
        +E_{4}^{A(2)}\left(u_{a} u_{c} s_{b} s_{d}+u_{a} u_{d} s_{b} s_{c}+u_{b} u_{c} s_{a} s_{d}+u_{b} u_{d} s_{a} s_{c}\right) .
    \end{aligned} 
\end{equation}
Contracting the expression with $\delta^{a b} \delta^{c d}$, $\left(\delta^{a c} \delta^{b d}+\delta^{b c} \delta^{a d}\right)$,$\dots$, \\ $\left(u^{a} u^{c} s^{b} s^{d}+u^{a} u^{d} s^{b} s^{c}+u^{b} u^{c} s^{a} s^{d}+u^{b} u^{d} s^{a} s^{c}\right)$, 
the linear system of equations for the coefficients $A^{A(2)}$, $B^{A(2)}$, $\dots$, $E_{4}^{A(2)}$ is obtained, 
which can be expressed as matrix equations as 
\begin{equation}
    M^{(2)} X^{A(2)} = Y^{A(2)}.
\end{equation}
Here 
$$
M^{(2)}=
 \left[\begin{array}{cccccccccccc}
    9 & 6 & 6 & 6 & 4 & 4 & 1 & 1 & 2 & 4 \eta^{2} \\
    6 & 24 & 4 & 4 & 16 & 16 & 2 & 2 & 4 \eta^{2} & 4\left(1+\eta^{2}\right) \\
    6 & 4 & 8 & 2+6 \eta^{2} & 8 & 8 \eta^{2} & 2 & 2 \eta^{2} & 2\left(1+\eta^{2}\right) & 8 \eta^{2} \\
    6 & 4 & 2+6 \eta^{2} & 8 & 8 \eta^{2} & 8 & 2 \eta^{2} & 2 & 2\left(1+\eta^{2}\right) & 8 \eta^{2} \\
    4 & 16 & 8 & 8 \eta^{2} & 24 & 4+20 \eta^{2} & 4 & 4 \eta^{2} & 8 \eta^{2} & 4+12 \eta^{2} \\ 
    4 & 16 & 8 \eta^{2} & 8 & 4+20 \eta^{2} & 24 & 4 \eta^{2} & 4 & 8 \eta^{2} & 4+12 \eta^{2} \\ 
    1 & 2 & 2 & 2 \eta^{2} & 4 & 4 \eta^{2} & 1 & \eta^{4} & 2 \eta^{2} & 4 \eta^{2} \\ 
    1 & 2 & 2 \eta^{2} & 2 & 4 \eta^{2} & 4 & \eta^{4} & 1 & 2 \eta^{2} & 4 \eta^{2} \\ 
    2 & 4 \eta^{2} & 2\left(1+\eta^{2}\right) & 2\left(1+\eta^{2}\right) & 8 \eta^{2} & 8 \eta^{2} & 2 \eta^{2} & 2 \eta^{2} & 2\left(1+\eta^{4}\right) & 8 \eta^{2} \\ 
    4 \eta^{2} & 4\left(1+\eta^{2}\right) & 8 \eta^{2} & 8 \eta^{2} & 4+12 \eta^{2} & 4+12 \eta^{2} & 4 \eta^{2} & 4\eta^{2} & 8 \eta^{2} & 4\left(1+\eta^{2}\right)^{2}\end{array}\right] ,
$$
$$
X^{A(2)} = \left[A^{A(2)}, B^{A(2)},C_{1}^{A(2)},C_{2}^{A(2)},
D_{1}^{A(2)},D_{2}^{A(2)},E_{1}^{A(2)},E_{2}^{A(2)},E_{3}^{A(2)},E_{4}^{A(2)}\right]^{-1} ,
$$
where $\eta=\hat{u} \cdot \hat{s}$.
And the coefficients in $Y^{A(2)}$ can be calculated by contracting eq.~(\ref{gamma2}) with $\delta^{a b} \delta^{c d}$, $\left(\delta^{a c} \delta^{b d}+\delta^{b c} \delta^{a d}\right)$,$\dots$
For example, \\ $Y^{T(2)}(1) = -\frac{1}{6} \int d^{2} \Omega_{\hat{n}}\left(\frac{\beta}{2}[1+\hat{n} \cdot \hat{u}]\right)^{2} \sum_{A=+,\times} e_{a b}^{A} e_{c d}^{A}\delta^{a b} \delta^{c d} e^{-i \alpha \hat{n} \cdot \hat{s}}=0$.
The solution of coefficient $X^{A(2)}$ is
\begin{equation}
    X^{A(2)}=(M^{(2)})^{-1}Y^{A(2)} .
\end{equation}
The coefficients $Y^{A(2)}$ for tensor, vector, scalar-breathing and scalar-longitudinal modes are 
$$
Y^{T(2)}=\frac{4\pi\beta^2}{3}
\left[\begin{array}{c}
0 \\
-j_0(\alpha)+(2i\eta-1/\alpha)j_1(\alpha)+\eta^2j_2(\alpha) \\
0 \\
0 \\
\left(\frac{3}{\alpha ^2}+\frac{4 i \eta }{\alpha }+1\right) j_2(\alpha )+\eta ^4 j_4(\alpha )+\left(-\frac{6 \eta
   ^2}{\alpha }+2 i \eta  \left(\eta ^2-1\right)\right) j_3(\alpha )-\frac{3 j_1(\alpha )}{\alpha }\\
\left(\frac{2 \left(\eta ^2-2\right)}{5 \alpha ^2}+\frac{4 i \eta }{\alpha }\right) j_2(\alpha )+\frac{2 \eta ^2
   j_3(\alpha )}{\alpha }-\frac{2 j_1(\alpha )}{\alpha } \\
\left\{
\begin{aligned}
    & \left(-\frac{8}{\alpha^3}+\frac{2 i \eta }{\alpha ^2}-\frac{\eta ^2}{\alpha }\right) j_3(\alpha )
    +\left(\frac{11}{\alpha ^2}+\frac{2 i \eta  \left(\eta ^2-1\right)}{\alpha }\right) j_4(\alpha ) \\
    &+\left(\frac{-\eta ^4+4 \eta ^2-5}{2 \alpha }+\frac{1}{4} i \eta  \left(\eta^2-1\right)^2\right) j_5(\alpha )
    +\frac{1}{8} \left(\eta ^2+1\right)\left(\eta ^2-1\right)^2 j_6(\alpha )
\end{aligned} \right\}\\
\frac{\left(\eta ^2+1\right) j_4(\alpha )}{\alpha ^2}+\left(\frac{2 \left(\eta ^2-5\right)}{\alpha ^3}+\frac{2 i \eta
}{\alpha ^2}\right) j_3(\alpha ) \\
\frac{\left(15 \eta ^4-6 \eta ^2-1\right) j_4(\alpha )}{2 \alpha ^2}+\left(\frac{8-24 \eta ^2}{\alpha ^3}+\frac{2 i \eta 
\left(5 \eta ^2-3\right)}{\alpha ^2}\right) j_3(\alpha ) \\
\left(\frac{2 \left(3 \eta ^4-6 \eta ^2+7\right)}{\alpha ^2}+\frac{2 i \eta  \left(\eta ^2-1\right)}{\alpha }\right)
j_4(\alpha )+\left(\frac{-6 \eta ^2-26}{\alpha ^3}+\frac{2 i \left(3 \eta ^3+\eta \right)}{\alpha ^2}\right)
j_3(\alpha )+\frac{\left(\eta ^4-1\right) j_5(\alpha )}{\alpha }
\end{array}\right] ,
$$

$$
Y^{V(2)}=\frac{4\pi\beta^2}{3}
\left[\begin{array}{c}
0 \\
\eta ^2 j_2(\alpha )+\left(2 i \eta -\frac{1}{\alpha }\right) j_1(\alpha )-j_0(\alpha )\\
0 \\
0 \\
\left(-\frac{14}{\alpha ^2}+\frac{8 i \eta }{\alpha }\right) j_2(\alpha )
-\frac{1}{2} \left(\eta ^2+1\right)^2 j_4(\alpha)
+\left(\frac{10 \eta ^2+6}{\alpha }-i \left(\eta ^3+\eta \right)\right) j_3(\alpha ) \\
\left(\frac{-\eta ^2-13}{\alpha ^2}+\frac{8 i \eta }{\alpha }\right) j_2(\alpha )
+\left(\frac{6 \eta ^2+10}{\alpha }-2 i\eta \right) j_3(\alpha )+\left(-\eta ^2-1\right) j_4(\alpha ) \\
\left\{\begin{aligned}
    &\left(-\frac{10}{\alpha ^3}+\frac{6 i \eta }{\alpha ^2}\right) j_3(\alpha )
    +\left(\frac{9 \eta ^2+7}{\alpha ^2}+\frac{i \eta  \left(\eta ^2-3\right)}{\alpha }\right) j_4(\alpha) \\
    &+\left(\frac{15 \eta ^4-18 \eta ^2-1}{2\alpha }-i \eta ^3 \left(\eta ^2-1\right)\right)j_5(\alpha )
    +\frac{1}{2} \left(\eta ^2-\eta ^6\right)j_6(\alpha )
\end{aligned} \right\}\\
\left(\frac{-\eta^2-9}{\alpha ^3}+\frac{6 i \eta }{\alpha ^2}\right) j_3(\alpha )
+\left(\frac{4 \left(\eta ^2+3\right)}{\alpha ^2}-\frac{2 i \eta }{\alpha }\right) j_4(\alpha )
+\frac{\left(-\eta ^2-1\right) j_5(\alpha )}{\alpha }\\
\left(\frac{10-30 \eta ^2}{\alpha ^3}+\frac{4 i \eta  \left(4 \eta ^2-1\right)}{\alpha ^2}\right)j_3(\alpha )
+\left(\frac{15\eta ^2\left(\eta ^2+2\right)-13}{\alpha ^2}-\frac{4 i \eta  \left(2 \eta ^2-1\right)}{\alpha }\right)j_4(\alpha )
+\frac{\left(1-5 \eta ^4\right) j_5(\alpha )}{\alpha } \\
\left\{\begin{aligned}
    &\left(\frac{-11 \eta ^2-29}{\alpha ^3}+\frac{4 i \eta  \left(\eta ^2+5\right)}{\alpha ^2}\right) j_3(\alpha)
    +\left(\frac{-3 \eta ^4+4 \eta ^2-17}{2 \alpha }-i \eta  \left(\eta^2-1\right)\right) j_5(\alpha )\\
    &+\left(\frac{3\eta ^4+35 \eta ^2+26}{\alpha ^2}-\frac{i \eta  \left(3 \eta ^2+5\right)}{\alpha }\right) j_4(\alpha) 
    +\frac{1}{2} \left(1-\eta ^4\right) j_6(\alpha )
\end{aligned} \right\}
\end{array}\right] ,
$$

$$
Y^{B(2)}=\frac{4\pi\beta^2}{3}
\left[\begin{array}{c}
\left(\eta ^2+1\right) j_2(\alpha )+\left(2 i \eta -\frac{4}{\alpha } \right) j_1(\alpha )\\
\left(\eta ^2+1\right) j_2(\alpha )+\left(2 i \eta -\frac{4}{\alpha } \right) j_1(\alpha ) \\
\left(-\frac{12}{\alpha ^2}+\frac{4 i \eta }{\alpha }\right) j_2(\alpha )
+\left(\eta^4-1\right) j_4(\alpha )
+\left(\frac{10-6 \eta ^2}{\alpha }+2 i \eta  \left(\eta ^2-1\right)\right) j_3(\alpha) \\
\frac{2 \left(\eta ^2+1\right) j_3(\alpha )}{\alpha }
+\left(\frac{2 \left(\eta ^2-7\right)}{\alpha ^2}+\frac{4 i \eta}{\alpha }\right) j_2(\alpha ) \\
\left(-\frac{12}{\alpha ^2}+\frac{4 i \eta }{\alpha }\right) j_2(\alpha )
+\left(\eta ^4-1\right) j_4(\alpha)
+\left(\frac{10-6 \eta ^2}{\alpha }+2 i \eta  \left(\eta ^2-1\right)\right) j_3(\alpha )\\
\frac{2 \left(\eta ^2+1\right)}{\alpha }j_3(\alpha )
+\left(\frac{2 \left(\eta ^2-7\right)}{\alpha ^2}+\frac{4 i \eta}{\alpha }\right) j_2(\alpha ) \\
\left\{\begin{aligned}
    &\left(-\frac{16}{\alpha ^3}+\frac{4 i \eta }{\alpha ^2}\right) j_3(\alpha )
    +\left(\frac{22-18 \eta ^2}{\alpha ^2}+\frac{4 i \eta  \left(\eta ^2-1\right)}{\alpha }\right) j_4(\alpha) \\
    &+\frac{1}{4} \left(\eta ^2+1\right)\left(\eta ^2-1\right)^2 j_6(\alpha )
    +\left(\frac{-\eta ^4+6 \eta ^2-5}{\alpha }+\frac{1}{2} i \eta  \left(\eta^2-1\right)^2\right) j_5(\alpha )
\end{aligned} \right\}\\
\frac{2 \left(\eta ^2+1\right)}{\alpha ^2}j_4(\alpha )
+\left(\frac{4 \left(\eta ^2-5\right)}{\alpha ^3}+\frac{4 i \eta}{\alpha ^2}\right) j_3(\alpha )\\
\left(\frac{2 \left(3 \eta ^4-6 \eta ^2+7\right)}{\alpha ^2}+\frac{2 i \eta  \left(\eta ^2-1\right)}{\alpha }\right)j_4(\alpha )
+\left(\frac{2 i \left(3 \eta ^3+\eta \right)}{\alpha ^2}-\frac{6 \eta ^2+26}{\alpha ^3}\right)j_3(\alpha )
+\frac{\left(\eta ^4-1\right)}{\alpha }j_5(\alpha )\\
\left(\frac{21 \eta ^4-18 \eta ^2+13}{\alpha ^2}+\frac{2 i \eta  \left(\eta ^2-1\right)}{\alpha }\right) j_4(\alpha)
+\left(-\frac{2 \left(27 \eta ^2+5\right)}{\alpha ^3}+\frac{2 i \eta  \left(13 \eta ^2-5\right)}{\alpha ^2}\right)j_3(\alpha )
+\frac{\left(\eta ^4-1\right)}{\alpha }j_5(\alpha )
\end{array}\right] 
$$
and 
$$
Y^{L(2)}=\frac{4\pi\beta^2}{3}
\left[\begin{array}{c}
\frac{1}{2} \left(\eta ^2+1\right) j_2(\alpha )+\left(i \eta  -\frac{2}{\alpha }\right) j_1(\alpha ) \\
\left(\eta ^2+1\right) j_2(\alpha )+\left(2 i \eta -\frac{4}{\alpha } \right) j_1(\alpha )\\
\left(-\frac{8}{\alpha ^2}+\frac{6 i \eta }{\alpha }\right) j_2(\alpha )- \left(\eta ^2+1\right) \eta ^2
j_4(\alpha )+\left(\frac{13 \eta ^2+1}{ \alpha }-2i \eta ^3\right) j_3(\alpha )\\
\left(\frac{-2\eta ^2-6}{\alpha ^2}+\frac{6 i \eta }{\alpha }\right) j_2(\alpha )+\left(\frac{5 \eta ^2+9}{ \alpha }-2i
\eta \right) j_3(\alpha )- \left(\eta ^2+1\right) j_4(\alpha )\\
\left(-\frac{16}{\alpha ^2}+\frac{12 i \eta }{\alpha }\right) j_2(\alpha )-2\left(\eta ^2+1\right) \eta ^2 j_4(\alpha
)+2\left(\frac{13 \eta ^2+1}{\alpha }-2 i \eta ^3\right) j_3(\alpha )\\
\left(-\frac{4 \left(\eta ^2+3\right)}{\alpha ^2}+\frac{12 i \eta }{\alpha }\right) j_2(\alpha )+2\left(\frac{5 \eta
^2+9}{\alpha }-2 i \eta \right) j_3(\alpha )-2\left(\eta ^2+1\right) j_4(\alpha )\\
\left\{\begin{aligned}
    &\left(-\frac{18}{\alpha ^3}+\frac{15 i \eta }{\alpha ^2}\right) j_3(\alpha )
    +\left(\frac{3 \left(33 \eta ^2+1\right)}{2 \alpha ^2}-\frac{10 i \eta ^3}{\alpha }\right) j_4(\alpha)\\
    &+\frac{1}{2} \left(\eta ^6+\eta^4\right) j_6(\alpha )
   +\left(-\frac{\eta ^2 \left(13 \eta ^2+3\right)}{\alpha }+i \eta ^5\right)
   j_5(\alpha )
\end{aligned} \right\}\\
\left\{\begin{aligned}
    &\left(\frac{3 \left(13 \eta ^2+21\right)}{2 \alpha ^2}-\frac{10 i \eta }{\alpha }\right) j_4(\alpha )
    +\left(-\frac{6\left(\eta ^2+2\right)}{\alpha ^3}+\frac{30 i \eta }{2 \alpha ^2}\right) j_3(\alpha ) \\
    &+\left(-\frac{7 \eta ^2+9}{\alpha }+i \eta \right) j_5(\alpha )
    +\frac{1}{2} \left(\eta ^2+1\right) j_6(\alpha )
\end{aligned} \right\}\\
\left\{\begin{aligned}
    &2\left(-\frac{13 \eta ^2+5}{\alpha ^3}+\frac{3 i \eta  \left(2 \eta ^2+3\right)}{\alpha^2}\right) j_3(\alpha )
    +\left(-\frac{9 \eta ^4+22 \eta ^2+1}{\alpha }+2i \eta ^3\right) j_5(\alpha )\\
    &+\left(\frac{12 \eta ^4+77 \eta ^2+13}{\alpha ^2}
    -\frac{2i \eta  \left(7 \eta ^2+3\right)}{\alpha}\right) j_4(\alpha )
   + \left(\eta ^4+\eta ^2\right) j_6(\alpha )
    \end{aligned} \right\}\\
\left\{\begin{aligned}
    &\left(-\frac{4 \left(13 \eta ^2+5\right)}{\alpha ^3}+\frac{12 i \eta  \left(2 \eta ^2+3\right)}{\alpha ^2}\right)j_3(\alpha )
    +2\left(-\frac{9 \eta ^4+22 \eta ^2+1}{\alpha }+2 i \eta^3\right) j_5(\alpha )\\
    &+2\left(\frac{12 \eta ^4+77 \eta ^2+13}{\alpha ^2}-\frac{2 i \eta  \left(7 \eta ^2+3\right)}{\alpha }\right) j_4(\alpha)
   +2\left(\eta ^4+\eta ^2\right) j_6(\alpha )
    \end{aligned} \right\}
\end{array}\right]
$$
respectively.

Combining eq.~(\ref{Gamma_TDI}), eq.~(\ref{gamma2}) and eq.~(\ref{gamma2_con}), we get the ORF frequency second-order expansion,
\begin{equation}
    \begin{aligned} 
        \Gamma_{IJ}^{A2}(&f)=\frac{e^{\frac{i}{2}(\beta_J-\beta_I)}}{32 \pi} \sum_{i,  j } P_{I, i} P_{J, j }
        \bigg\{ A^{A(0)}+A^{A(2)}_I+A^{A(2)}_J 
        +2\left(B^{A(0)}+B^{A(2)}_I+B_J^{A(2)}\right)\left(\hat{u}_{I,i} \cdot \hat{u}_{J,j}\right)^2 \\ 
        &+\left(C^{A(0)}+C_{I,1}^{A(2)}+C_{J,1}^{A(2)}\right)
        \left(\left(\hat{u}_{I,i} \cdot \hat{s}_{i j}\right)^{2}+\left(\hat{u}_{J,j} \cdot \hat{s}_{i j}\right)^{2}\right)
        +\left(C_{I,2}^{A(2)}+C_{J,2}^{A(2)}\right)\left(\left(\hat{u}_{I,i} \cdot \hat{u}_{J,j}\right)^{2}+1\right) \\ 
        &+4\left(D^{A(0)}+D_{I,1}^{A(2)}+D_{J,1}^{A(2)}\right)\left(\hat{u}_{I,i} \cdot \hat{u}_{J,j}\right)
        \left(\hat{u}_{I,i} \cdot \hat{s}_{i j}\right)\left(\hat{u}_{J,j} \cdot \hat{s}_{i j}\right)
        +4\left(D_{I,2}^{A(2)}+D_{J,2}^{A(2)}\right)\left(\hat{u}_{I,i} \cdot \hat{u}_{J,j}\right)^{2} \\ 
        &+\left(E^{A(0)}+E_{I,1}^{A(2)}+E_{J,1}^{A(2)}\right)
        \left(\hat{u}_{I,i} \cdot \hat{s}_{i j}\right)^{2}\left(\hat{u}_{J,j} \cdot \hat{s}_{i j}\right)^{2}
        +\left(E_{I,2}^{A(2)}+E_{J,2}^{A(2)}\right)\left(\hat{u}_{I,i} \cdot \hat{u}_{J,j}\right)^{2} \\ 
        &+E_{I,3}^{A(2)}\left(\left(\hat{u}^{\prime}_{j } \cdot \hat{s}_{i j}\right)^{2}
        +\left(\hat{u}_{I,i} \cdot \hat{s}_{i j}\right)^{2}\left(\hat{u}_{I,i} \cdot \hat{u}_{J,j}\right)^{2}\right)
        +E_{J,3}^{A(2)}\left(\left(\hat{u}_{I,i} \cdot \hat{s}_{i j}\right)^{2}
        +\left(\hat{u}_{J,j} \cdot \hat{s}_{i j}\right)^{2}\left(\hat{u}_{I,i} \cdot \hat{u}_{J,j}\right)^{2}\right) \\ 
        &+4\left(E_{I,4}^{A(2)}+E_{J,4}^{A(2)}\right)\left(\hat{u}_{I,i} \cdot \hat{u}_{J,j}\right)
        \left(\hat{u}_{I,i} \cdot \hat{s}_{i j}\right)\left(\hat{u}_{J,j} \cdot \hat{s}_{i j}\right)\bigg\} .
    \end{aligned} 
\end{equation}

\acknowledgments
This work is supported by the National Key R\&D Program of China under Grant No. 2022YFC2204602, the Natural Science Foundation of China (Grants No. 12247154, No. 11925503), the Postdoctoral Science Foundation of China (Grant No. 2022M711259), Guangdong Major project of Basic and Applied Basic Research (Grant No. 2019B030302001).


\end{document}